\documentclass{article}
\pdfoutput=1

\usepackage{arxiv}

\usepackage[utf8]{inputenc} 
\usepackage[T1]{fontenc}    
\usepackage{hyperref}       
\usepackage{url}            
\usepackage{booktabs}       
\usepackage{amsfonts}       
\usepackage{nicefrac}       
\usepackage{microtype}      
\usepackage{lipsum}
\usepackage{graphicx}
\usepackage{multicol}
\usepackage{graphicx} 
\usepackage{subcaption} 
\usepackage{float}
\usepackage{amsmath}
\graphicspath{ {./images/} }

\usepackage{hyperref}
\hypersetup{
    colorlinks=true,       
    linkcolor=black,       
    citecolor=black,       
    filecolor=black,       
    urlcolor=black         
}

\usepackage[square,numbers]{natbib}
\title{Integrating Agent-Based and Compartmental Models for Infectious Disease Modeling: A Novel Hybrid Approach}

\author{
 Inan Bostanci \\
  Zuse Institute Berlin\\
  14195 Berlin \\
  \texttt{bostanci@zib.de} \\
   \And
 Tim Conrad \\
    Zuse Institute Berlin\\
    14195 Berlin \\
  \texttt{conrad@zib.de} \\
}

\begin{document}
\maketitle
\begin{abstract}
This study investigates the spatial integration of agent-based models (ABMs) and compartmental models for infectious disease modeling, presenting a novel hybrid approach and examining its implications.
ABMs offer detailed insights by simulating interactions and decisions among individuals but are computationally expensive for large populations.
Compartmental models capture population-level dynamics more efficiently but lack granular detail.
We developed a hybrid model that aims to balance the granularity of ABMs with the computational efficiency of compartmental models, offering a more nuanced understanding of disease spread in diverse scenarios, including large populations.
This model spatially couples discrete and continuous populations by integrating an ordinary differential equation model with a spatially explicit ABM. Our key objectives were to systematically assess the consistency of disease dynamics and the computational efficiency across various configurations. For this, we evaluated two experimental scenarios and varied the influence of each sub-model via spatial distribution. In the first, the ABM component modeled a homogeneous population; in the second, it simulated a heterogeneous population with landscape-driven movement. Results show that the hybrid model can significantly reduce computational costs but is sensitive to between-model differences, highlighting the importance of model equivalence in hybrid approaches. The code is available at: \url{git.zib.de/ibostanc/hybrid_abm_ode}
\end{abstract}

\keywords{Hybrid Model \and Coupled Models \and Agent-Based Model \and Compartmental Model \and Epidemiology \and Infectious Disease Modeling}

\section{Introduction}
Preventing or mitigating the widespread transmission of infectious diseases is a critical objective for health systems worldwide. Disease modeling is essential to develop effective intervention strategies and analyze their societal impacts, as underscored by the COVID-19 pandemic \citep{lancet_comission_how_2024}. In this context, agent-based models (ABMs) and compartmental models are fundamental approaches \citep{krause_modeling_2019}. As a result, these models have gained significant recognition in epidemiology and are invaluable to policymakers and public health officials \citep{becker_development_2021}. 

Compartmental models are traditionally used for mathematical modeling of infectious diseases and capture high-level system dynamics by describing the rates of transition between sub-populations, such as susceptible, infected, and recovered \citep{hethcote_mathematics_2000}. They offer valuable insights on spreading dynamics at the macro-scale, but their reliance on a homogeneously mixed population limits realism \citep{brauer_mathematical_2019, duan_mathematical_2015, wulkow_prediction_2021}. Although meta-population models can increase their complexity and spatial resolution by further disaggregating the population (e.g. through mobility \cite[][]{chang_mobility_2021, calvetti_metapopulation_2020, winkelmann_mathematical_2021} or layered populations \cite[][]{gao_epidemic_2022}), they retain the simplifying assumption of homogeneity within each sub-population.  

In contrast, ABMs simulate the behavior of individual agents, accounting for their interactions, movements, and decisions. This enables a detailed analysis of disease transmission under various scenarios, providing a more granular and realistic approach to infectious disease modeling \citep{railsback_agent-based_2019, bruch_agent-based_2015}. ABMs allow for population heterogeneity, for example, by representing the population as a network of agents \citep{kerr_covasim_2021}, or by capturing complex spatial-temporal patterns through explicit modeling of movement and interactions \citep{arifin_landscape_2015, muller_predicting_2021, merler_spatiotemporal_2015}. This makes ABMs valuable for exploring individual-level variability and allows for a straightforward implementation of policy interventions.

However, this increased realism comes at a cost: ABMs often require more complex, fine-grained data for fitting and verification \citep{krause_modeling_2019}. While their emergent macro-level outcomes can be validated with observed data, their implemented micro-level relationships are often based on unobserved assumptions \citep{brailsford_hybrid_2019, frias-martinez_agent-based_2011}, making them less adept to derive underlying parameters than compartmental models, such as the basic reproduction number \citep{mokhtari_multi-method_2021}. Further, ABMs require significantly more computational resources compared to compartmental models, a discrepancy that grows with population size \citep{ozmen_analyzing_2016, hunter_comparison_2018}.

These challenges hinder ABM-based analysis of large populations and call for a hybrid modeling approach that combines the computational and analytical efficiency of compartmental models with the detailed insights of ABMs.
Such a hybrid model could integrate multiple geographic regions with different population characteristics and data availability, while allowing for population movement between regions. Moreover, it may scale up to larger populations without sacrificing significant predictive power. 

Despite their potential, hybrid models have received relatively little attention in infectious disease modeling research. While the meta-population approach couples sub-populations under the same modeling paradigm and retains the assumption of homogeneity within these, hybrid approaches could flexibly loosen or constrain assumptions for different populations where appropriate, harnessing the strengths and data requirements of both model types.
Few studies have explored such hybrid models, most of which have not critically examined how model coupling impacts the dynamics of infectious diseases under varying conditions. 

To fill this gap, we propose a hybrid model that spatially integrates an ABM with a compartmental model based on ordinary differential equations (ODEs). One of the primary strengths of this study is its methodical assessment of the effects of spatial coupling on key indicators that reflect the infectious disease dynamics. 
Further, we specifically address challenges in coupling discrete and continuous population representations under different modeling paradigms and systematically explore the feasibility of the hybrid approach by addressing the following: (1) How does spatial coupling of an ABM with an ODE-based model affect infection spread dynamics? (2) What are the computational cost savings compared to a pure ABM? 

To answer these, we study two scenarios in a simplified toy example:
In the first, random movement in the ABM component results in a homogeneous population. In the second, landscape-influenced movement encourages spatial heterogeneity, and therefore deviates from the ODE component. 
In both scenarios, we systematically alter the spatial distribution and the interface location between sub-models and examine the resulting impact on key indicators for infection dynamics and computational efficiency. Our results show that the spatial location of the coupling mechanism can affect the peak time and shape of the infectious curve, particularly when movement patterns differ across regions. These findings underscore the importance of understanding the effects of hybrid model coupling for accurate epidemic modeling.

The article is structured as follows: We begin by reviewing related work on hybrid models that couple compartmental models with ABMs, particularly in the field of infectious disease modeling. Next, we detail the implementation of the ABM, the compartmental model, and the hybrid model, with an explicit description of the coupling mechanism. Following, we vary the spatial distribution of model components within the hybrid model in two experiments to assess the consistency of infection dynamics across configurations. 
Finally, we discuss the results and give an outlook for future research.

\section{Related Work}
One of the earliest hybrid model approaches for infectious diseases is described by \citet{bobashev_hybrid_2007}. This study implemented a temporal coupling that alternates the model between an ABM state and a compartmental model state, with the switch being triggered when the number of infected individuals crosses a threshold value. Aggregate population values of the ABM were used as initial conditions for the compartmental model state. The authors argued that insights into epidemic processes at the local level diminish in informative value with an increasing number of infected individuals. They evaluated several threshold values and demonstrated how such a model can be applied to a single hypothetical city. Further, using air travel data, they extended this model to a network of cities. \citep{hunter_hybrid_2020} followed a similar approach, applying the switching method to a meta-population setting for several regions in Ireland. They compared computational execution times and highlighted the potential of this model to conserve resources. However, they also noted that the computationally most efficient configurations might yield inferior model results due to the suboptimal performance of the compartmental model component, measured by the length and final size of the outbreak.
Rather than using infection numbers, \citet{vinh_toward_2016} implemented a hybrid model that triggers a state switch when local agent density crosses a threshold. The authors argued that dense clusters of agents can be viewed as a homogeneous population, and therefore a compartmental model is sufficient.

Another approach involves using agents to couple compartmental models in a hybrid model. For example, \citet{nguyen_hybrid_2022} investigated the role of temporary staff during the COVID-19 pandemic by modeling care homes as compartmental models and temporary staff as agents moving between them.
Similarly, in \citet{bradhurst_improving_2016}, agents represent herds of livestock, and within each agent, ODEs determine infection dynamics. Agent-agent interactions govern the spread of diseases between herds through different pathways. In \citet{banos_importance_2015}, cities are represented as network nodes with infectious disease dynamics modeled by compartmental models. A hybrid model, where agent-based air travel connects nodes with possible transmission during flight, was compared to a meta-population approach. The hybrid model showed slower infection spread for scenarios with few infected agents per node due to the discrete, stochastic mobility implementation, contrasting the even diffusion in the deterministic meta-population model.
A similar coupling mechanism was employed in \citet{marilleau_coupling_2018} to model the spread of a vole population in a French region. This spatial model divides the region into grid cells, allowing vole agents to travel between those cells. Within each cell, the vole population size is determined by equation-based models. Once the population of a cell reaches a critical value, the cell generates vole agents that leave the cell and migrate to another cell with more suitable conditions.

A problem that arises when coupling an ABM with a compartmental model is the differing population shape in each model type. In an ABM, the population is represented by distinct entities, while an ODE model represents the population as continuous matter \citep{gustafsson_guide_2016}. \citet{sewall_interactive_2011} highlight that this is particularly problematic when agents leave a compartmentally modeled region due to information loss when converting agents to a continuous population. 
In \citet{solano_coupling_2013}, this caused a total population decline because the agent-population size was fixed while compartmental population sizes were unconstrained, leading to asymmetric population flow. Additionally, individuals entering the ABM from a compartmental node were represented in aggregated floating point numbers, and reconstructing these into integer agent counts lead to discrepancies. 
Addressing this disparity is crucial in developing a successful hybrid model, as it involves accurately transitioning the population between discrete entities and continuous mass representations.

In a review of hybrid models in operational research, \citet{brailsford_hybrid_2019} note that published studies often have limited reporting on verification and validation, with many failing to sufficiently describe the architecture of the hybrid models used. Although their review focused on operational research, we found similar issues in the hybrid infectious disease models reviewed in this section. Specifically, these models frequently omit details on interactions between components and the mechanisms of data exchange between sub-models. Additionally, as with the studies in \citet{brailsford_hybrid_2019}, the implications of these hybrid models for infectious disease prediction and control are often underexplored, a notable gap given the importance of model accuracy in public health contexts.

In conclusion, while various hybrid models have been developed to leverage the strengths of both ABMs and compartmental models, they often face challenges such as determining appropriate switching mechanisms and managing the transition between discrete and continuous representations of the population. In the following section, we describe our proposed hybrid model and how we addressed these challenges effectively.

\section{The Hybrid Model}
This section introduces our proposed hybrid modeling approach, which combines the strengths of a microscopic ABM with a macroscopic compartmental ODE model. The hybrid approach aims to balance computational efficiency with a detailed and realistic representation of infection dynamics. In the hybrid model, infectious disease dynamics are modeled by two parallel components: the first simulates infections via agent-agent interactions in a continuous coordinate space; the second employs ODEs to model these dynamics. This design leverages the fine-grained interactions and spatial detail of ABMs while utilizing the computational simplicity of compartmental models to scale up to larger populations effectively.

The following sections detail the agents, environment, and coupling mechanism, illustrating how our hybrid model addresses the challenges of combining discrete and continuous population representations. Moreover, we demonstrate the model's potential to reduce computational costs while maintaining accurate and insightful predictions of infectious disease dynamics.
\subsection{Sub-Model M1: Agent-Based Model (ABM)}
\label{M1}
The ABM simulates agents within a 2D rectangular region in a xy-plane, defined by the coordinates $0 \leq x \leq 9$ and $0 \leq y \leq 9$. Each agent occupies a $(x,y)$-coordinate within this space and is in one of three states: susceptible, infectious, or recovered. The probability $p_i$ of a susceptible agent $i$ to become infectious within a time step is $p_i = 1 - (1-\rho)^k$, 
where $\rho$ is the base probability and $k$ is the number of infectious agents within a distance of $\theta$ to the susceptible agent. This transmission model assumes a fixed transmission probability upon contact and increases with repeated exposure, akin to a contact-duration model \cite[see][]{frias-martinez_agent-based_2011, breitwieser_biodynamo_2022}. While dose-response models are more realistic \citep{jones_dose-response_2015, aganovic_new_2023}, we chose to implement this simplified model due to its similarity to the ODE's transmission mechanism: In the ODE model, the transmission rate is a function of the contact rate and the infection probability. In our ABM, the contact rate can be seen as a function of the radius size and the step size. Infectious agents recover at a rate of $1/D$ per time step, where $D$ is the average time an agent remains in the infectious state, becoming immune to subsequent infections through agent-agent interactions.

\subsubsection{Agent Movement}
Agent movement is implemented in two variants. In variant M1-A, agents move randomly, simulating a homogeneous population. During a time step, an agent can move in any direction with a maximum possible step size of $\Lambda$ under periodic boundary conditions. Although less realistic, this resembles the population in a compartmental model more closely. 

In variant M1-B, however, we implemented landscape-driven agent movement \cite[see][]{djurdjevac_conrad_human_2018, marshall_abmanimalmovement_2022}.
This landscape induces non-random movement and is generated by a bivariate density function
\begin{equation}
f(x,y) = -e^{-\frac{(x - X)^2 + (y - Y)^2}{\eta}} 
\label{eq:2}
\end{equation}
In this setup, agents either move towards (attraction) or away from (repulsion) lower elevation points. Here, the valley has its center at the coordinates $(X, Y)$ and the width of the valley is determined by the parameter $\eta$. Specific rules govern the transition between attraction and repulsion states, allowing for regular movement patterns across the simulation space. This leads to spatial heterogeneity of the agent population. 

\begin{figure}[h]
\begin{center}
    \includegraphics[scale=0.5]{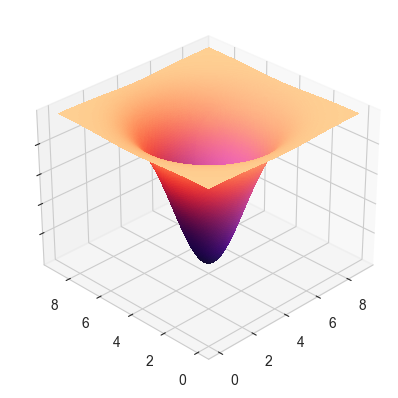}
    \caption{Visualization of the bivariate density function with X = Y = 4.5.}
    \label{fig:landscape}
  \end{center}
\end{figure}

\begin{figure}[h]
\begin{center}
    \includegraphics[scale=0.2]{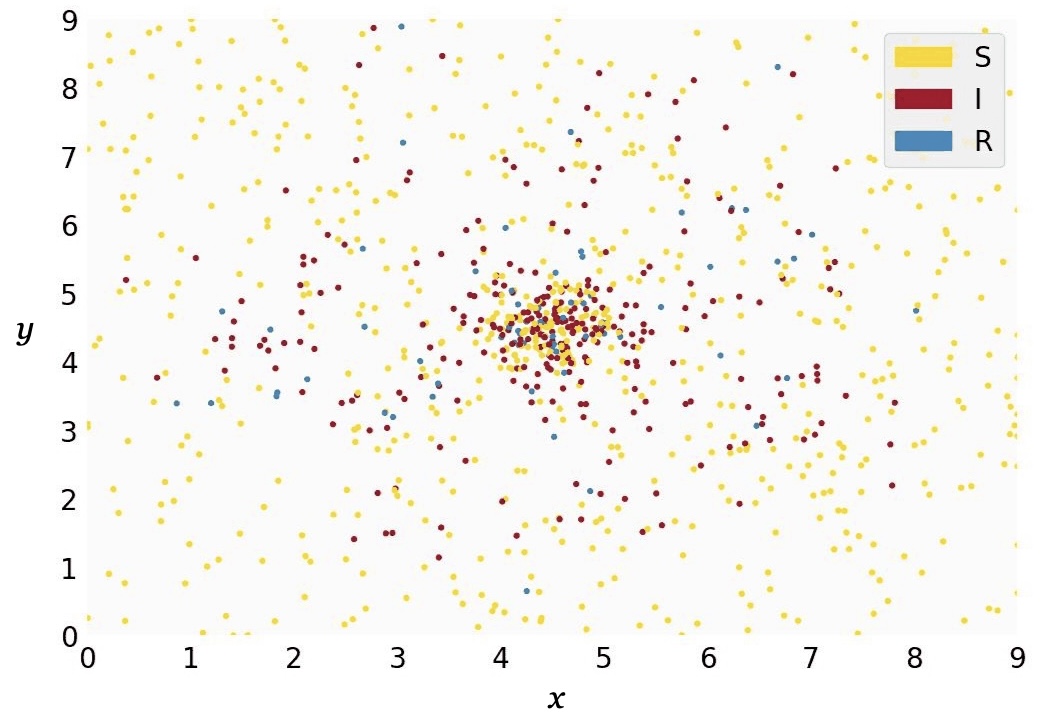}
    \caption[Snapshot]{Snapshot of an ABM realization (variant M1-B) at $t=9$. Each point represents the position of a susceptible (yellow), infectious (red), or recovered (blue) agent.}
    \label{fig:snapshot_abm}
    \end{center}
\end{figure}

At each time step, an agent evaluates the gradient of its current position on the xy-plane based on the underlying density function with

\begin{equation}
\chi_{i,t} = \frac{f(x_{i,t} + \delta, y_{i,t}) - f(x_{i,t} - \delta, y_{i,t})}{2\delta} - \lambda_{x,i,t}
\label{eq:3}
\end{equation}
\begin{equation}
\psi_{i,t} = \frac{f(x_{i,t}, y_{i,t} + \delta) - f(x_{i,t}, y_{i,t} - \delta)}{2\delta} - \lambda_{y,i,t}
\label{eq:4}
\end{equation}

where $\delta$ is a fixed base step size and $\lambda$ is a random value between $-\Lambda$ and $\Lambda$, introducing stochasticity in agent-movement. In the attraction state, the agent's new coordinates are

\begin{equation}
x_{_{i,t+1}}, y_{_{i,t+1}} = x_{_{i,t}} + \chi_{_{i,t}}, y_{_{i,t}} + \psi_{_{i,t}}
\end{equation}

and in the repulsion state, they are 

\begin{equation}
x_{_{i,t+1}}, y_{_{i,t+1}} = x_{_{i,t}} - \chi_{_{i,t}}, y_{_{i,t}} - \psi_{_{i,t}}
\end{equation}

under periodic boundary conditions. Agents switch to the repulsion state once they approached the center of the valley within a distance of 0.3 and switch back to attraction at a rate of $1/(3\eta)$. This dynamic switching mechanism allows for movement across the simulation space, resembling commute-like behavior, instead of the entire population remaining around the center for the entire simulation time. A snapshot of an M1-B realization at $t=9$ can be found in Figure \ref{fig:snapshot_abm}, showing a dense cluster of agents at the center, surrounded by less densely populated space.

\subsubsection{Initialization}
For initialization, agents are placed randomly across the simulation space in M1-A. In M1-B, initial positions and behavior states are determined using the Monte Carlo method, based on the bivariate density function for the coordinates and $1/(1-1/3\eta)$ for the behavior state probabilities, with the outbreak starting after a burn-in period.

\subsection{Sub-Model M2: Compartmental Model}
\label{compartmental_model}

The compartmental model follows the classic SIR (Susceptible-Infectious-Recovered) framework, describing the dynamics of disease spread in a population with ODEs \citep{kermack_contribution_1927, brauer_mathematical_2019}. The model divides the population into three compartments: susceptible individuals ($S$), infectious individuals ($I$), and recovered individuals ($R$). The rates at which the population transitions between compartments are given by:

\begin{equation}
\begin{aligned}
\frac{{dS}}{{dt}} &= -\frac{{\beta \cdot S \cdot I}}{{N}} \\
\frac{{dI}}{{dt}} &= \frac{{\beta \cdot S \cdot I}}{{N}} - \gamma \cdot I \\
\frac{{dR}}{{dt}} &= \gamma \cdot I
\end{aligned}
\end{equation}

where $\beta$ represents the transmission rate of the disease, $\gamma$ is the recovery rate, and $N$ is the total population size. Given initial conditions, step size, and time frame, the ODEs can be solved numerically, providing compartment sizes over time - and thus simulating the dynamics of disease spread in a population.

\subsection{Coupling}
\label{coupling}

In the hybrid model, the population-space is spatially divided into an ABM (M1) and a compartmental region (M2), while allowing for population flow between regions. The interface between M1 and M2 is a vertical line that can be moved along the x-axis to adjust the proportions (or weight) of each sub-model. This flexible interface allows us to vary the influence of each model within the hybrid framework.

The temporal resolution and the order of progresses can significantly impact the results of both ABMs and compartmental models \citep{ozmen_analyzing_2016}. This aspect is particularly crucial in our hybrid model due to the interaction between a discrete model and a continuous model \citep{gustafsson_guide_2016}. As we implemented discrete time steps in the ABM, we maintain coherence by iteratively solving the ODE model for one time step while interactions within the ABM and between sub-models occur synchronously. 

\begin{figure*}[t]
    \begin{center}
    \includegraphics[width=\textwidth]{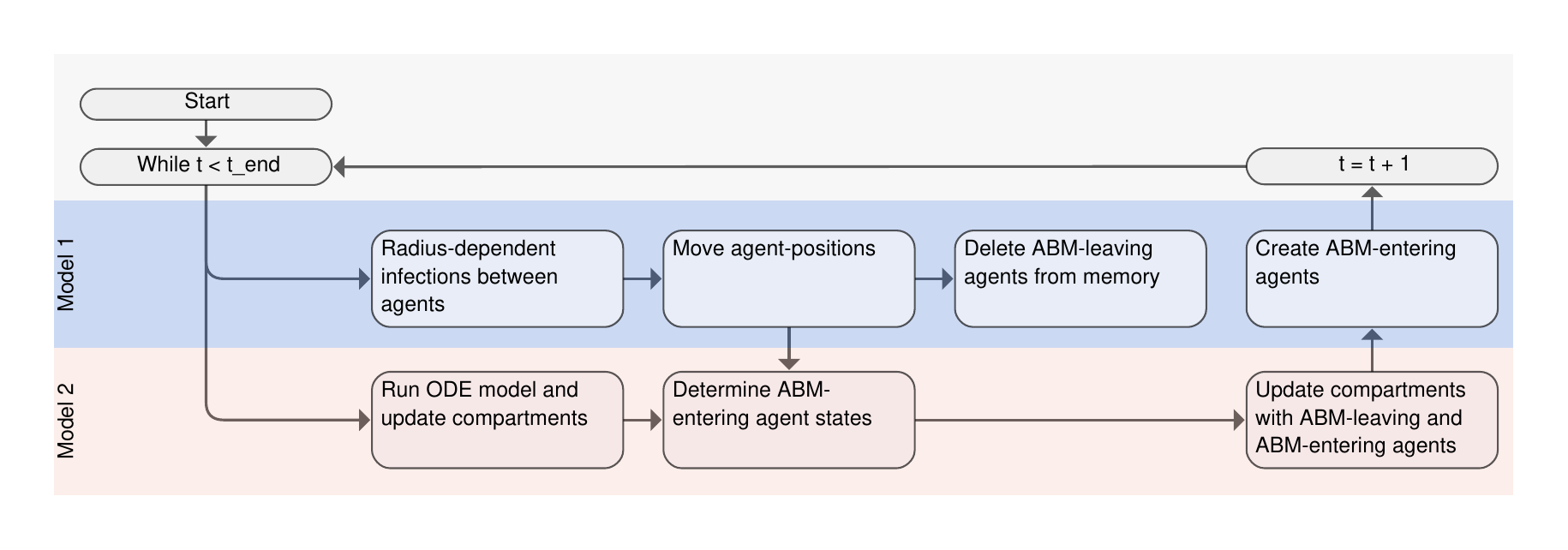}
    \caption[Flowchart]{Order of processes in hybrid model.}
    \label{fig:flowchart}
    \end{center}
\end{figure*}

As discussed earlier, the exchange between discrete and continuous population representations requires careful handling to ensure accurate population tracking and infection dynamics. Figure \ref{fig:flowchart} visualizes the coupling mechanism in the hybrid model, depicting the sequence of processes within a single time step. The subsequent sections detail these processes.

\subsubsection{Population Flow between Model Regions} 
Agents can transition from the M1 (ABM) area to the M2 (ODE) area through the designated interface. Agents entering M2 are aggregated into $s_{in}$, $i_{in}$, and $r_{in}$, representing the number of susceptible, infectious, and recovered agents respectively. These agents are subsequently removed from the M1 population. Movement from M2 into M1 in a given time step is governed by the equation:

\begin{equation}
n_{out}(t) = s_{in}(t-1) + i_{in}(t-1) + r_{in}(t-1)
\end{equation}

where $n_{out}$ is the number of agents leaving M2. This ensures that the flow between the two regions remains symmetric. The lag of one time step prevents scenarios where more agents leave M2 than are present, thus avoiding population inconsistencies. Since M2 does not simulate spatial movement, the $y$-coordinate at which an agent leaves M2 is unknown. For M1-A, this coordinate is sampled from a uniform distribution. For M1-B, the marginal of the density function (see Equation \ref{eq:2}) is used to derive sampling weights.

\subsubsection{Infection Dynamics within the ODE Area}

Due to population flow between models, the population size of M2 varies based on the incoming and outgoing agents. Additionally, the stochastic nature of the hybrid model requires a corrective procedure in M2 to ensure accurate population tracking and infection dynamics. This involves several steps:

First, the compartmental model described in Section \ref{compartmental_model} is solved for one time step with the current population size as shown in the following equations:

\begin{equation}
\begin{aligned}
\Sigma_1(t+1) &= S(t) - \frac{\beta \cdot S(t) \cdot I(t)}{N(t)}dt \\
\Phi_1(t+1) &= I(t) + \frac{\beta \cdot S(t) \cdot I(t)}{N(t)}dt - \gamma \cdot I(t)dt \\
\Psi_1(t+1) &= R(t) + \gamma \cdot I(t)dt
\end{aligned}
\end{equation}

Next, the state of the leaving population is determined. The state of each leaving agent is drawn in a multinomial trial using the results from the previous step as the class probabilities:

\begin{equation}
s_{out}(t),i_{out}(t), r_{out}(t) \sim \textrm{Multinomial}\left(n_{out}(t),
\left[\frac{\Sigma_1(t+1)}{N(t)}, \frac{\Phi_1(t+1)}{N(t)}, \frac{\Psi_1(t+1)}{N(t)}\right]\right)
\end{equation}

The leaving population is then subtracted from each compartment, ensuring no negative sizes by constraining each value to zero or higher (for conciseness, we will  demonstrate the following procedures for the S-compartment, but they also applied to the I- and R-compartment):

\begin{equation}
\begin{aligned}
\Sigma_2(t+1) &= \textrm{max}(0, \Sigma_1(t+1) - s_{out}(t)) 
\end{aligned}
\end{equation}

Given this reduction, the sum of the compartments may exceed the actual remaining population, calculated as $N(t) - n_{out}(t)$. Therefore, a correction term $\kappa$ is computed to balance this discrepancy:

\begin{equation}
\begin{aligned}
P(t+1) &= \Sigma_2(t+1) + \Phi_2(t+1) + \Psi_2(t+1) \\
\kappa(t) &= P(t+1) - (N(t) - n_{out}(t))
\end{aligned}
\end{equation}

This correction term is proportionally subtracted from each compartment to ensure that their sum matches the actual population:

\begin{equation}
\begin{aligned}
\Sigma_3(t+1) &= \Sigma_2(t+1) - \kappa(t) \cdot \frac{\Sigma_2(t+1)}{P(t+1)} 
\end{aligned}
\end{equation}

As a final step, the entering population is added to the compartments, resulting in the final compartment sizes for the following time step: 

\begin{equation}
\begin{aligned}
S(t+1) &= \Sigma_3(t+1) + s_{in}(t) \\
\end{aligned}
\end{equation}

\subsubsection{Parameter Selection}
To effectively compare and couple M1 and M2, both models need to produce similar epidemic outcomes given the same starting conditions. However, an ABM featuring a heterogeneous population, as is the case with M1-B, cannot produce identical results to an analogous compartmental model \citep{susandi_relation_2021}. Our aim with M1-B is therefore to achieve the closest feasible alignment with the M2 outcomes.

Producing comparable dynamics can be achieved by parametrizing one model first and adjusting the other model to match. 
\citet{rahmandad_heterogeneity_2008} argue that in the real world, where contact structures are often unknown, compartmental model parameters are usually fitted to epidemic data, effectively capturing the impact of network topology and individual variability. 
However, with known values for the basic reproduction number ($R_0$) and disease duration, parametrizing a compartmental model first and fitting the ABM parameters afterwards simplifies calibration and requires fewer assumptions. We thus adopted this latter approach and followed the methodology and disease characteristics for a measles epidemic outlined in \citet{breitwieser_biodynamo_2022}, allowing us to focus on evaluating the hybrid coupling mechanism without introducing additional complexity. 

Using parameters $\beta=1.61$ and $\gamma=1/8$ for a measles epidemic with $R_0=12.9$ and an 8-day recovery period, we generated a time series of infectious individuals with M2. This was done by numerically solving the compartmental model with initial values of $S=995, I=5, R=0$ and a step size of 0.33 days, using the RK45 method as implemented in SciPy's integrate.solve\_ivp function \citep{virtanen_scipy_2020}.

We then fitted the parameters $\Lambda$, $\rho$, and $\theta$ for M1-A to match the time series data, with the same initial conditions and a recovery rate of $1/24$ (corresponding to a step size of 0.33 days). This optimization was done using the Nelder-Mead minimizer (as implemented in SciPy's optimize.minimize \citep{virtanen_scipy_2020}) with least squares as the cost function, yielding $\Lambda= 0.215$, $\rho=0.025$, and $\theta=1.162$. 

We repeated the procedure for M1-B, and additionally fitted the parameter $\delta$ (the base step size), fixing $\eta$ at 3, and positioning the attraction point at $X = Y = 4.5$. This results in $\Lambda= 0.268$, $\rho=0.016$, $\theta=1.01$, and $\delta=0.01$. 

\begin{figure}[H]
    \begin{center}
    \includegraphics[scale=0.6]{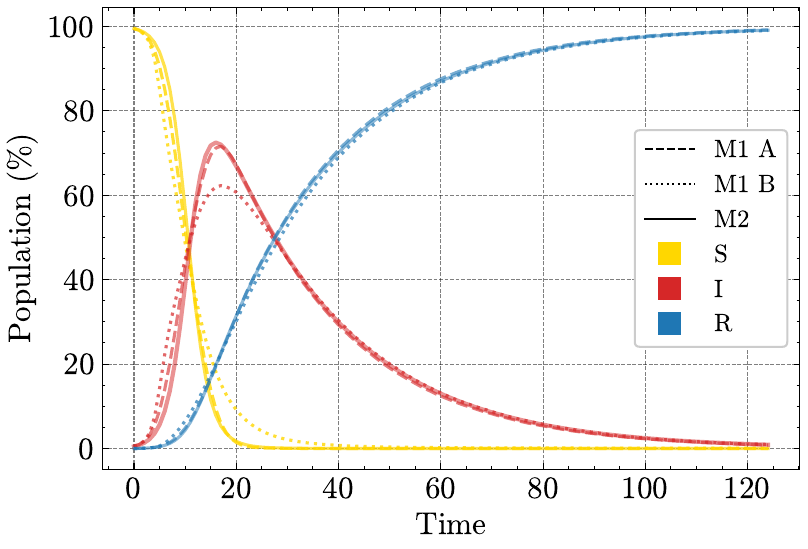}
    \caption[Separate SIR]{SIR curves for model 2 (solid lines) and the average results for 100 realizations of the fitted M1 variants A (dashed lines) and B (dotted lines).}
    \label{fig:fit}
    \end{center}
\end{figure}

Figure \ref{fig:fit} shows that M1-A and M2 produce closely aligned SIR curves, while M1-B exhibits a more rapid rise and a lower peak of infections. This discrepancy is expected because, under population heterogeneity, infections quickly spread in dense clusters. As a result, the overall pool of susceptible individuals decreases faster, leading to an accelerated initial rise and a reduced overall dispersion of infections \citep{rahmandad_heterogeneity_2008}.

In the following section, we describe the numerical experiments conducted to evaluate the performance and robustness of our hybrid modeling approach. The primary objective is to understand how varying the weight of the compartmental (ODE-based) and agent-based component affects model predictions, infection dynamics, and computational efficiency. By systematically adjusting these weights, we aim to uncover the nuanced behavior of the hybrid model under different configurations. This analysis provides valuable insights into the potential applications and limitations of the suggested hybrid model in infectious disease modeling.

\section{Experimental Setup and Methodology}
\label{experiment1}
In this section, we describe the experimental setup and methodology used to evaluate the fundamental dynamics of our hybrid model. By varying the weight between its two main components - the ABM (M1) and the compartmental model (M2) - we aim to understand how altering the balance between these components and the location of the interface affects model outcomes in terms of the infectious wave. This setup allows us to test the robustness of our model under various conditions. The results of these experiments are presented in the subsequent section.

\subsection{Model Weights} 
We adjusted the weights of M1 ($w_1$) and M2 ($w_2$) by shifting the interface in the hybrid model along the x-axis. For example, given $0 \leq x \leq 9$, placing the interface at x=4.5 results in $w_1 = w_2 =0.5$, while a placement at x=6.75 results in $w_1 = 0.25$ and $w_2=0.75$. The interface shift allows us to systematically study the impact of different model configurations on infection dynamics.

\subsection{Model Configurations} 
We compare two configurations of the hybrid model: Hybrid Model A (M1-A with M2) and Hybrid Model B (M1-B with M2), enabling us to examine its performance under two scenarios. In the first scenario, both model components of Hybrid Model A simulate a homogeneous population, leading to very similar results individually. In the second scenario, Hybrid Model B's agent-based component simulates a heterogeneous population, yielding distinct epidemic dynamics compared to the M2-component. In this scenario, the attraction point is placed at the center of the simulation space $(x=4.5, y=4.5)$, making the distance of the attraction point to the interface dependent on $w$. For $w_2>0.5$, the attraction point falls within the borders of M2. Additionally, we examine a third scenario in the appendix, where the the attraction point is always placed at the center of M1, rather than at the center of the full simulation space. These scenarios enable us to study the implications of spatial differences for spatial coupling. 

\subsection{Evaluation Metrics} 
For the experiments, we evaluate the following metrics for the resulting time series of the infectious population: (a) the root mean squared error (RMSE) compared to the baseline from M2, quantifying the overall deviation from the reference model; (b) the peak height; (c) the time step at which the peak occurs; (d) the Fisher-Pearson coefficient of skewness, indicating the buildup speed of the infectious wave; (e) the full-width-at-half-maximum (FWHM), indicating the sharpness of the peak. Metrics b - e provide detailed insights into the dynamics and shape of the epidemic. Finally, we measure (f) the average CPU time for each model configuration. We increase the weight of M2 from 0 to 1 in 0.02 steps, averaging the metrics after 100 independent realizations per configuration.

\section{Results}

\begin{figure}[t]
 \begin{center}
  \begin{subfigure}[p]{0.33\textwidth}
    \includegraphics[width=\textwidth]{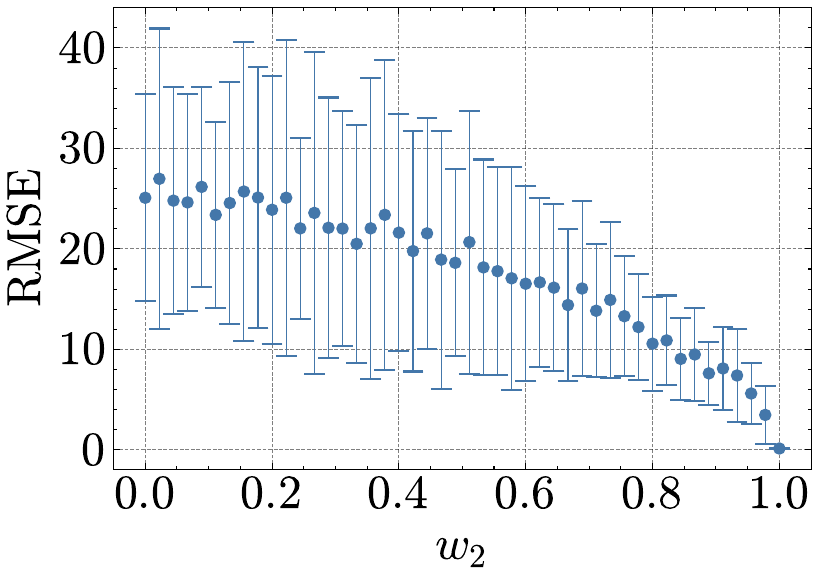}
    \caption{}
    \label{fig:rmsea}
  \end{subfigure}
  \begin{subfigure}[p]{0.33\textwidth}
    \includegraphics[width=\textwidth]{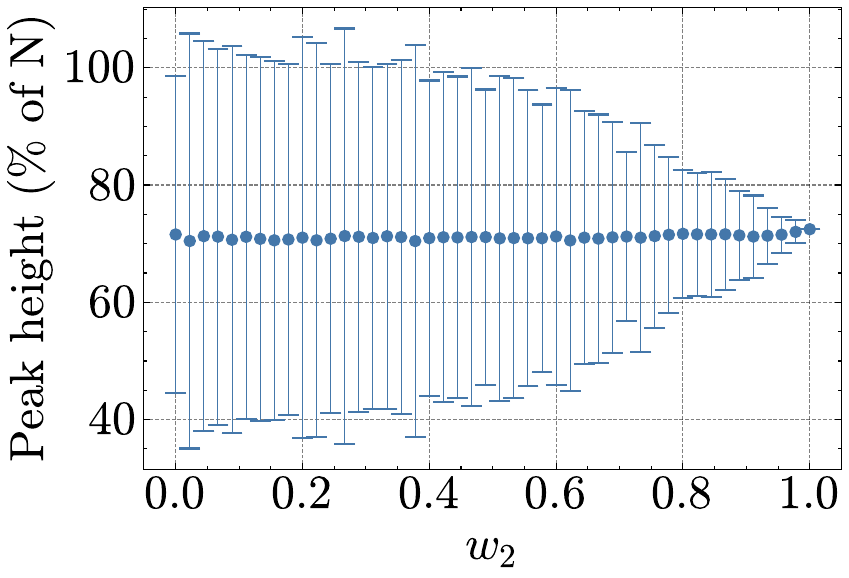}
    \caption{}
    \label{fig:max_infa}
  \end{subfigure}
  \begin{subfigure}[p]{0.33\textwidth}
    \includegraphics[width=\textwidth]{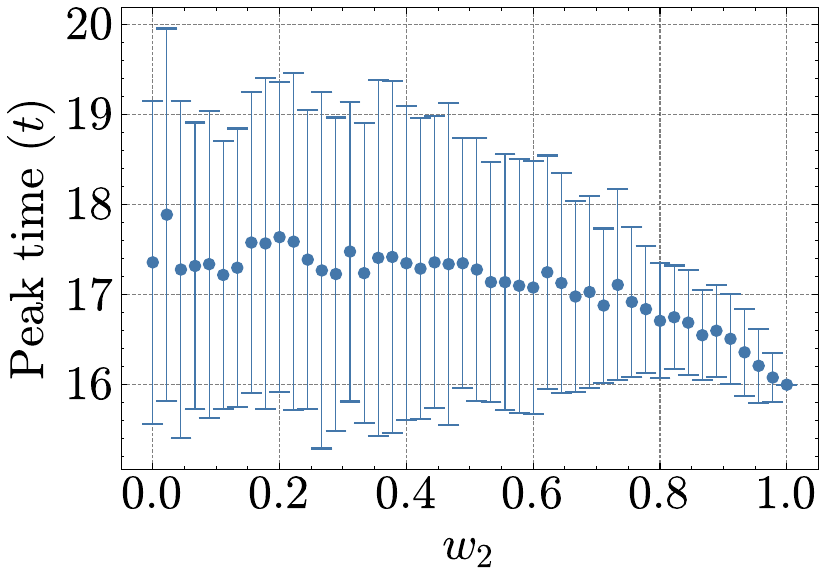}
    \caption{}
    \label{fig:t_peaka}
  \end{subfigure}
  \begin{subfigure}[p]{0.33\textwidth}
    \includegraphics[width=\textwidth]{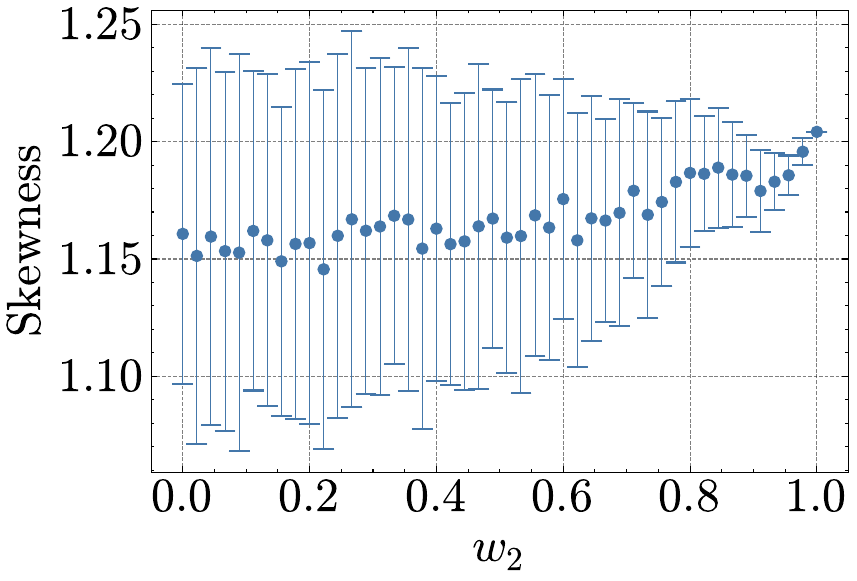}
    \caption{}
    \label{fig:skewnessa}
  \end{subfigure}
  \begin{subfigure}[p]{0.33\textwidth}
    \includegraphics[width=\textwidth]{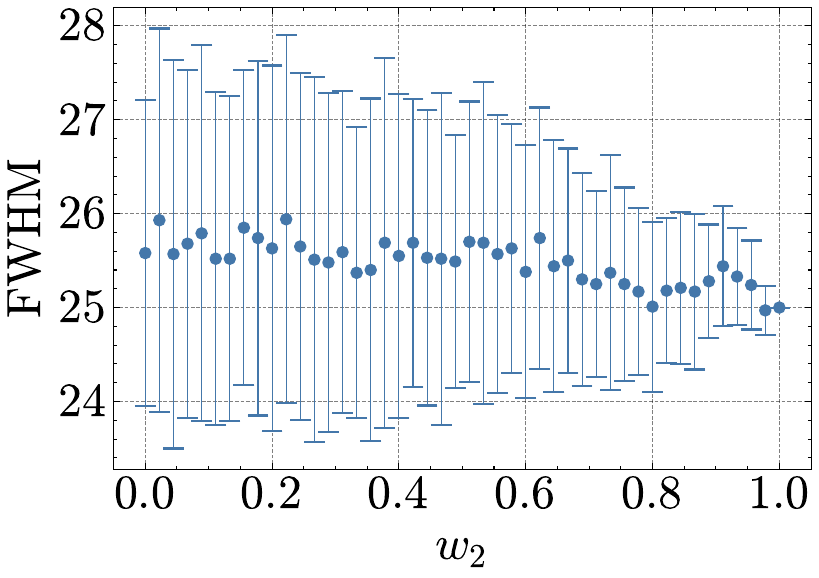}
    \caption{}
    \label{fig:fwhma}
  \end{subfigure}
  \begin{subfigure}[p]{0.33\textwidth}
    \includegraphics[width=\textwidth]{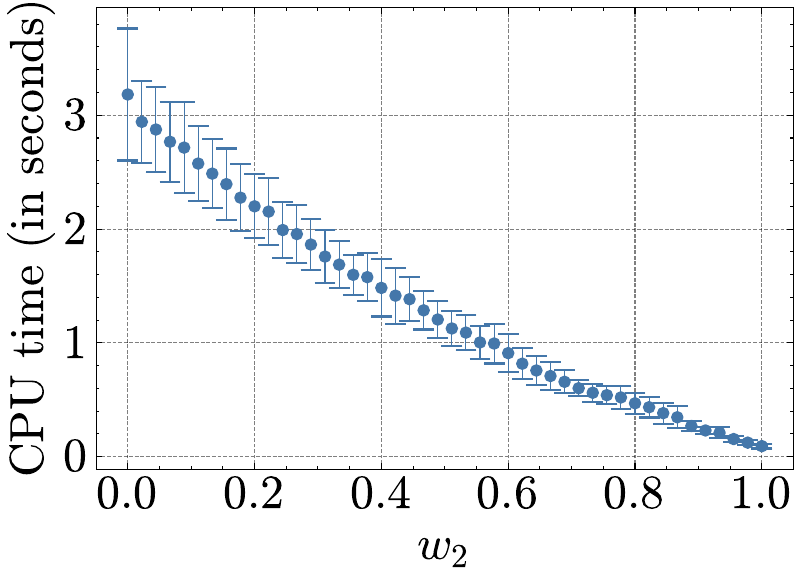}
    \caption{}
    \label{fig:cpu_timea}
  \end{subfigure}
   \end{center}
   \label{fig:hyba}
  \caption{Hybrid Model A: Average results for 100 independent realizations per M2 weight ($w_2$). Bars around points represent the standard deviation. (a) RMSE for the infectious time series between the reference model and the hybrid model. (b) Peak time of the infectious time series. (c) Peak height of the infectious time series. (d) Fisher-Pearson coefficient of skewness of the infectious time series. (e) FWHM of the infectious time series. (f) CPU time in seconds.}
\end{figure}

\begin{figure}[htbp]
 \begin{center}
  \begin{subfigure}[p]{0.33\textwidth}
    \includegraphics[width=\textwidth]{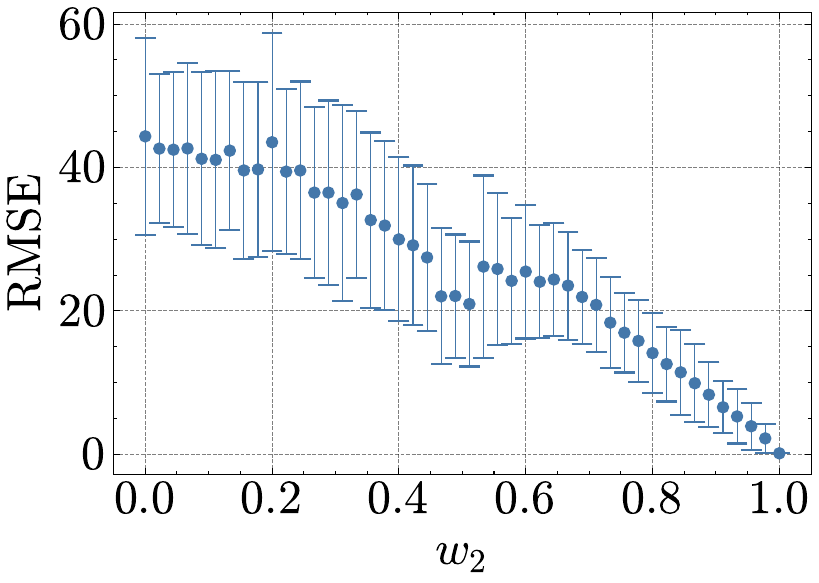}
    \caption{}
    \label{fig:rmseb}
  \end{subfigure}
  \begin{subfigure}[p]{0.33\textwidth}
    \includegraphics[width=\textwidth]{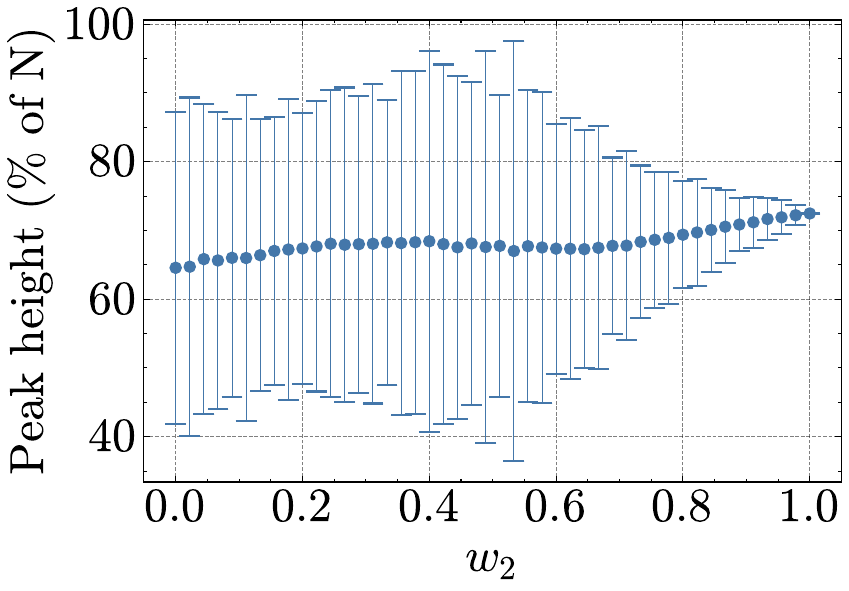}
    \caption{}
    \label{fig:max_infb}
  \end{subfigure}
  \begin{subfigure}[p]{0.33\textwidth}
    \includegraphics[width=\textwidth]{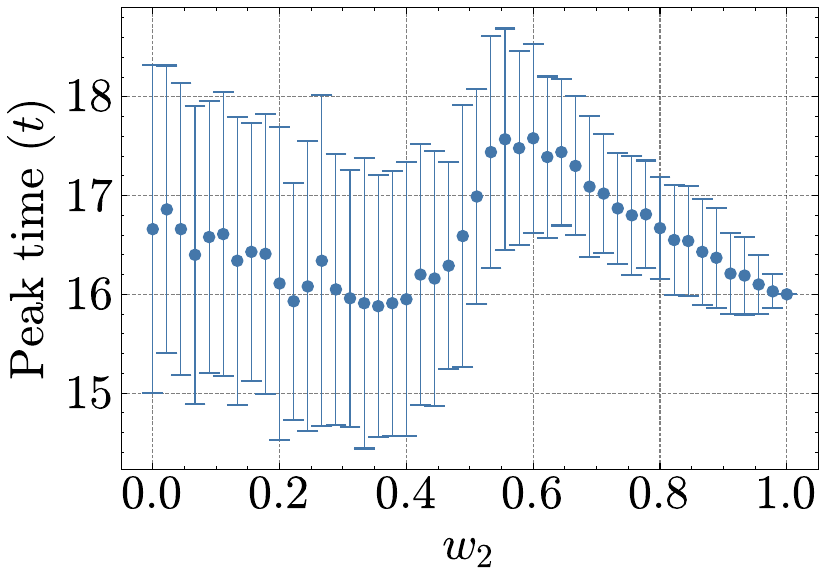}
    \caption{}
    \label{fig:t_peakb}
  \end{subfigure}
  \begin{subfigure}[p]{0.33\textwidth}
    \includegraphics[width=\textwidth]{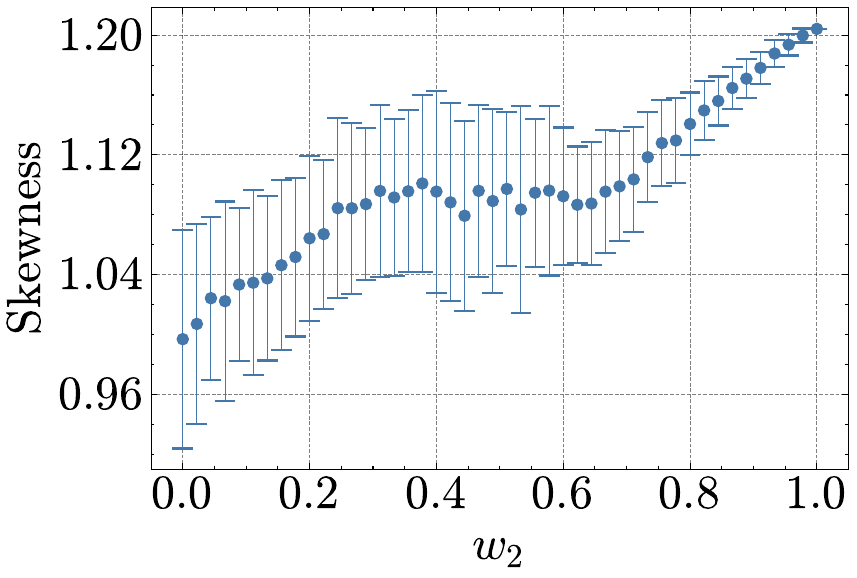}
    \caption{}
    \label{fig:skewnessb}
  \end{subfigure}
  \begin{subfigure}[p]{0.33\textwidth}
    \includegraphics[width=\textwidth]{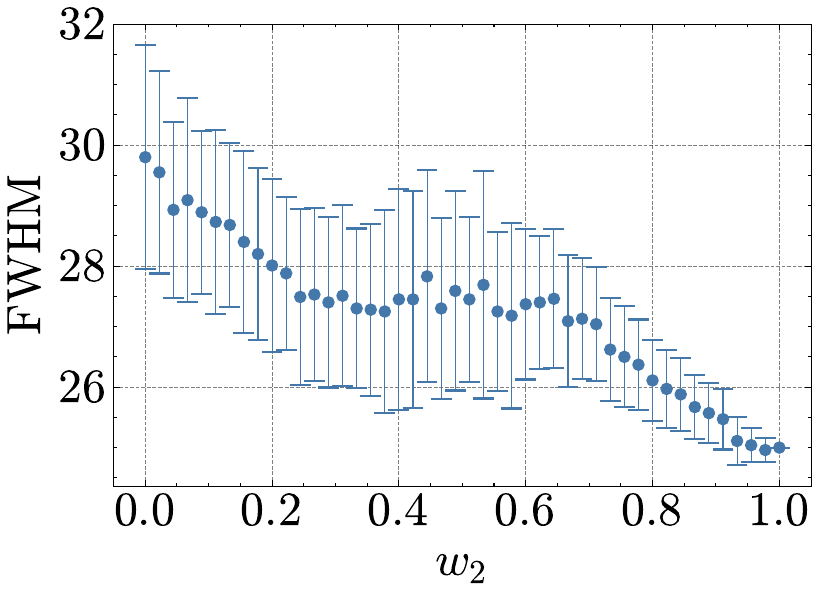}
    \caption{}
    \label{fig:fwhmb}
  \end{subfigure}
  \begin{subfigure}[p]{0.33\textwidth}
    \includegraphics[width=\textwidth]{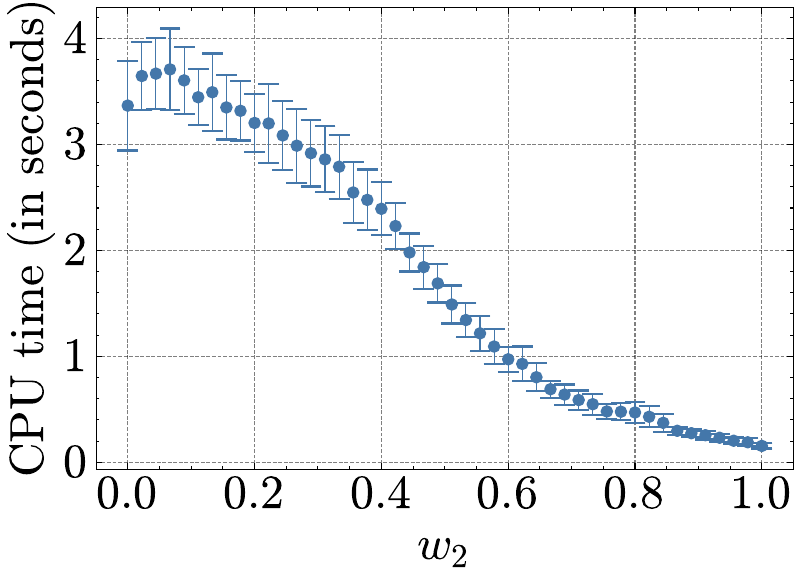}
    \caption{}
    \label{fig:cpu_timeb}
  \end{subfigure}
   \end{center}
   \label{fig:hybb}
  \caption{Hybrid Model B: Average results for 100 independent realizations per M2 weight ($w_2$). Bars around points represent the standard deviation. (a) RMSE for the infectious time series between the reference model and the hybrid model. (b) Peak time of the infectious time series. (c) Peak height of the infectious time series. (d) Fisher-Pearson coefficient of skewness of the infectious time series. (e) FWHM of the infectious time series. (f) CPU time in seconds.}
\end{figure}

This section presents the results of our experiments, evaluating the performance and robustness of the hybrid model under different configurations. Specifically, we examine how varying the weight of the ABM (M1) and the compartmental model (M2) affects various metrics (see previous section) related to infection dynamics and computational efficiency in both Hybrid Model A (M1-A with M2) and Hybrid Model B (M1-B with M2). Results are shown in Figures \ref{fig:rmsea}-\ref{fig:cpu_timea} for Hybrid Model A and in Figures \ref{fig:rmseb}-\ref{fig:cpu_timeb} for Hybrid Model B. The x-axis represents the weight of M2 ($w_2$). When $w_2=0$, the entire population is modeled by M1, and when $w_2=1$, the entire population is modeled by M2. 

\subsection{Hybrid Model A}
\subsubsection{Root Mean Squared Error (RMSE)} The RMSE for different weights behaves as expected, showing a steady drop from 25.09 to 0 as the weight of the M2-component increases. This is because the full M2-model serves as the reference model to compute the RMSE. The standard deviation is larger for lower values of $w_2$, which is due to the stochastic nature of M1. 

\subsubsection{Peak Height} The peak height of the infection curve does not change significantly across $w_2$-levels. It remains around 71.57\% at $w_2=0$ and 72.46\% at $w_2=1$. This consistency is expected since the peak height is similar in both individual models M1-A and M2, as shown in Figure \ref{fig:fit}. Similar to the RMSE, the standard deviation shrinks with increasing $w_2$. 

\subsubsection{Peak Time} A notable coupling effect is observed in the peak time of the infectious curve. While the peak occurs slightly earlier in the reference model ($16t$) compared to M1-A ($17.36t$), it does not gradually decrease as $w_2$ increases. Instead, there seems to be a quadratic relationship between the peak time and $w_2$. This may be due to the discrete implementation of time in our models and the relatively small difference in peak time of 1.36 time steps between the model components, causing the resulting peak time to align more closely with the model component modeling the majority of the population.

\subsubsection{Skewness} The skewness coefficient is 1.16 at $w_2=0$ and 1.2 at $w_2=1$, indicating slightly more weight on the right tail of the infectious wave in M2 compared to M1-A. The quadratic relationship in peak time is reflected in the skewness, which increases more rapidly as $w_2$ exceeds the 0.6-mark, likely due to the peak time influencing the skewness of the infectious wave.

\subsubsection{Full-Width-at-Half-Maximum (FWHM)} The FWHM is 25.58 at $w_2=0$ and 25 at $w_2=1$, indicating that the infectious wave of M2 is sharper compared to M1. However, this difference is minor and steadily decreases as $w_2$ increases, suggesting minimal coupling effects. 

\subsubsection{CPU Time} The CPU time decreases from 3.18 to 0.09 seconds as $w_2$ increases, following a convex pattern. This diminishing reduction in CPU time can be attributed to the stochastic processes computed for each agent, where calculations often involve interactions with all other agents. Thus, the reduction in computational load is more substantial when $w_2$ increases from lower values (e.g., 0 to 0.02) compared to higher values (e.g., 0.9 to 0.92), reflecting the non-linear scaling of interaction complexity.


\subsection{Hybrid Model B}

\subsubsection{Root Mean Squared Error (RMSE)} The RMSE at $w_2=0$ is 44.3, nearly twice as large as the RMSE for M1-A. Similar to Hybrid Model A, the RMSE for Hybrid Model B drops towards zero as $w_2$ increases. However, there is a notable minimum around the 0.5-mark of $w_2$. The RMSE decreases for $w_2$ values below 0.5, then increases for $w_2$ values above 0.5 before dropping again after the 0.6-mark. This indicates that M2 has a stronger effect on model results around the 0.5 mark than at the 0.6 mark. This can be attributed to the attraction point in the agent-based component, resulting in a spatially heterogeneous population: agents form a cluster where the infection spreads more rapidly. When $w_2$ is set to 0.5, the attraction point lies on the interface between M1-B and M2, attracting many agents into M2 and therefore increasing the interaction between the models. 

\begin{figure*}[htbp]
 \begin{center}
  \begin{subfigure}[p]{0.36\textwidth}
    \includegraphics[width=\textwidth]{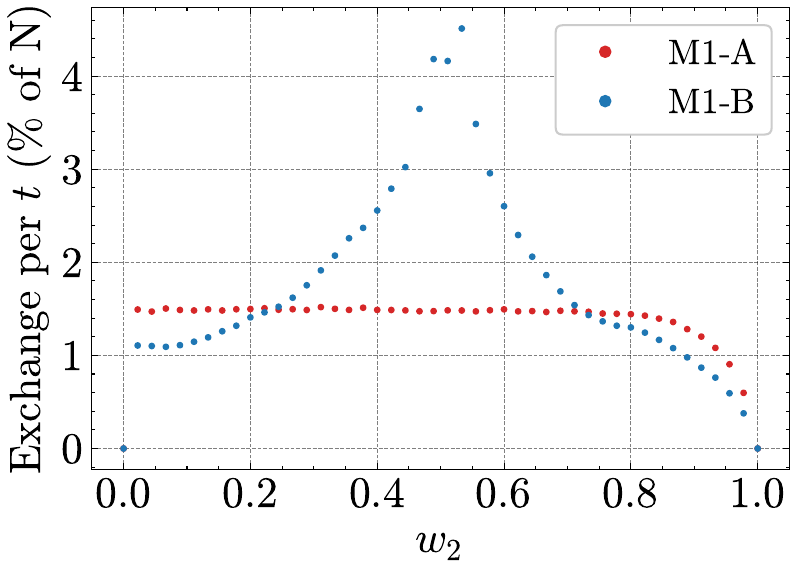}
    \caption{}
    \label{fig:mean_leave_abm}
  \end{subfigure}
  \begin{subfigure}[p]{0.38\textwidth}
    \includegraphics[width=\textwidth]{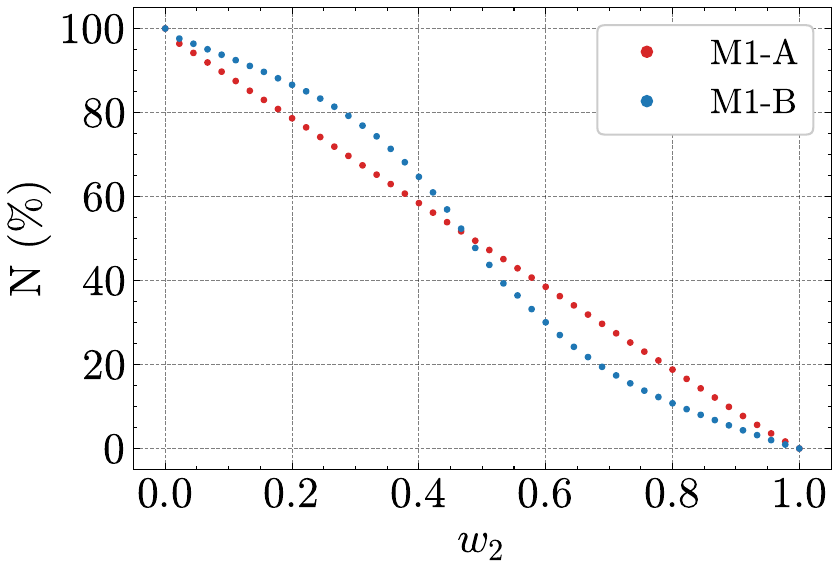}
    \caption{}
    \label{fig:mean_pop_abm}
  \end{subfigure}
   \end{center}
  \caption{Changes in population metrics for Hybrid Model A and B for different M2 weights ($w_2$). Results from 100 independent realizations. (a) Average percentage of population entering M2 per $t$. (b) Average percentage of total population in M1 per $t$.}
\end{figure*}

Indeed, Figure \ref{fig:mean_leave_abm} shows that the average population flow from M1 to M2 during a time step depends on $w_2$. In Hybrid Model A, it is around 1.5\% of the total population for weights below 0.8, before dropping for higher weights. In Hybrid Model B, it increases from 0.02 to 0.53, peaking at 4.5\%, before decreasing rapidly for higher weights. Additionally, Figure \ref{fig:mean_pop_abm} shows that the share of the total population present in M1 drops faster around $w=0.5$ in Hybrid Model B, while decreasing steadily in Hybrid Model A. 

\subsubsection{Peak Height} The peak height of M1-B is 64.56\%, considerably lower than in M2. A notable observation is the increased standard error around the 0.5-mark of $w_2$, reflecting the increased between-model interaction discussed earlier. The peak height follows a convex pattern: For $w_2 < 0.4$, the average peak height increases towards the higher value of M2, but decreases after the 0.4-mark until $w_2=0.6$. This pattern suggests that the attraction point's proximity to the interface significantly influences infection dynamics.

\subsubsection{Peak Time} The peak time of a full M1-B ($w_2=0$) is $16.66t$, having a delay of 0.66 that is slightly closer to the reference model than M1-A's peak. The attraction point appears to influence the peak time of the Hybrid Model B: The peak time is closer to the reference model around $w_2=0.4$ than around $w_2=0.6$, again likely due to the increased between-model interaction near the 0.5-mark. The local minimum near $w_2=0.4$ rather than around $w_2=0.5$ suggests that the discrete nature of time amplifies the interaction effect between the model components.

\subsubsection{Skewness}
The skewness of the resulting infectious data is considerably lower in M1-B than in M2, indicating a more symmetrical infection wave in M1-B. Although skewness increases with an increasing $w_2$, a local maximum at $w_2=0.38$ is followed by a minor drop until $w_2=0.64$. This reflects the shift observed in the previous metrics for Hybrid Model B.

\subsubsection{Full-Width-at-Half-Maximum (FWHM)} At $w_2=0$, the FWHM is considerably higher (29.8) than at $w_2=1$ (25.0), indicating that the peak in M1-B is more truncated than in M1-A or M2. Similar to the previous metric, a regression can be observed as $w_2$ increases with a temporary shift after the 0.38-mark. Both observations are consistent with the increased between-model exchange near this value. 

\subsubsection{CPU Time} For lower $w_2$-values, the average CPU time is higher for Hybrid Model B compared to Hybrid Model A, at 3.37 seconds when $w_2=0$. This is due to the increased complexity introduced by the attraction points. Additionally, there is an increase from $w_2=0$ to $w_2=0.06$, possibly due to the further complexity introduced by population flow from M2 to M1, influenced by the landscape in M1. For Hybrid Model B, the drop in CPU time follows the drop in population size of the M1-B component, as visualized in Figure \ref{fig:mean_pop_abm}. The computational demands of M1 stem from simulating each individual agent, making the population size a key factor in determining CPU time. 

\section{Discussion}

The results for Hybrid Model A demonstrate that the outcomes are directly proportional to the weighting of each model component. Specifically, as the weighting shifts, the model's outcomes increasingly resemble the results of the predominant component. This proportional relationship indicates that the hybrid model effectively balances the computational efficiency of the compartmental model (M2) with the detailed interactions of the agent-based model (M1-A). By adjusting the weight between these components, the hybrid model can be fine-tuned to optimize either efficiency or detail, providing a versatile and scalable approach to infectious disease modeling.

In contrast, the results for Hybrid Model B reveal a more complex interaction involving the landscape-driven agent movement and the between-model interface. This interaction triggers switches in the model outcomes, favoring one component over the other under different conditions. The direction-switching effect observed in the peak time and peak height for Hybrid Model B can be explained by the presence of the attraction point. Without the attraction point, these metrics would likely follow a trajectory similar to Hybrid Model A. However, the increased between-model interaction around $w_2=0.5$ leads to a local change, emphasizing the sensitivity of the hybrid model to the spatial configuration of the attraction point.

\subsection{Impact on Infection Spread Dynamics}
The study reveals that spatial coupling of an ABM with an ODE-based model can effectively capture the dynamics of infection spread in the overall population. In Hybrid Model A, where both components simulate a homogeneous population, the infection dynamics transition smoothly as the weight shifts between M1 and M2. This smooth transition reflects a successful integration of the detailed agent interactions in M1 with the aggregate population-level processes in M2, maintaining consistency in model outcomes.

In Hybrid Model B, where the agent-based component includes spatial heterogeneity, the infection dynamics reveal a deeper complexity. The presence of an attraction point introduces a non-uniform distribution of agents, leading to varied infection dynamics depending on the location of the interface. This demonstrates that spatial factors and agent clustering can significantly influence the spread of infection, underscoring the model's ability to adapt to complex scenarios by accurately reflecting the impact of spatial configurations and agent movements.

These findings highlight that the location of the coupling mechanism in a hybrid model can amplify the differences between model components. Differences between sub-models and sub-populations, as well as the level of population flow, should be carefully considered in real-world applications. The increased between-model interaction observed in Hybrid Model B underscores the need for a nuanced approach to spatial coupling, especially when dealing with heterogeneous populations, thereby enhancing the model's applicability and precision in various scenarios.

\subsection{Computational Efficiency}
Our experiments also provide insights into the computational efficiency of the hybrid model compared to a pure ABM. For Hybrid Model A, the hybrid approach achieves a substantial reduction in CPU time while maintaining accuracy in the infection dynamics. The proportional relationship between the weight of the compartmental model and the CPU time underscores the potential for significant computational savings. This is particularly beneficial in large-scale simulations where computational resources are a critical constraint.

In Hybrid Model B, while the overall trend shows reduced CPU time with an increasing weight of the compartmental model, the relationship is more complex. The initial increase in CPU time at low values of $w_2$ reflects the added complexity due to the attraction point and the increased interaction between sub-models. Despite this complexity, the hybrid model still demonstrates a net reduction in computational cost compared to a pure ABM, especially at higher values of $w_2$. This indicates that even with added spatial heterogeneity, the hybrid model can offer computational advantages, though the exact savings depend on the specific configuration and interaction dynamics. 

\subsection{Limitations and Future Directions}
One limitation of the hybrid model, compared to a pure ABM, is the loss of information during each exchange between the two sub-models. Whenever population flows from M2 to M1, the hybrid model creates new agents and randomly distributes their individual statuses based on M2's current state. Consequently, an agent that recently entered M2 as recovered may effectively leave as susceptible or infectious. Exploring alternative strategies for information exchange between sub-models and refining the coupling mechanism could potentially address this limitation and improve the accuracy of the hybrid model in capturing individual-level dynamics.

Overall, our findings provide valuable insights into the strengths and limitations of hybrid approaches. They emphasize the importance of carefully considering the spatial configuration and interaction dynamics between sub-models to optimize both accuracy and computational efficiency. However, real-world epidemic outbreaks often introduce incubation times, antibody decline, or age-dependent differences in susceptibility, infectiousness, and behavior, among others. While our focus on a simplistic susceptible-infected-recovered scenario allowed us to isolate and examine the effects of hybrid coupling, we recognize that it has inherent limitations in terms of real-world applicability. Nonetheless, the presented hybrid model could extend to epidemic models with more realistic compartment structures and corresponding additions in the ABM. As this requires that disease-specific parameters and interaction patterns are integrated with rigor, modelers need to carefully consider the trade-off between the development time of the hybrid model and its benefits in terms of cost savings and accuracy over a single model paradigm approach.

Further, our toy model does not take full advantage of computationally more efficient ABM frameworks, making generalizations in terms of computational cost savings difficult. Given the experimental nature of our study, we limited our experiments to a small scale, focusing on the coupling mechanism and its implications for disease dynamics. In this case, modeling a larger population with our toy model would likely favor the hybrid approach even more, as the computational demands of the ODE component remain constant regardless of population size. We aimed to balance this potential discrepancy by modeling a relatively small population. 

Future research could develop more sophisticated coupling mechanisms and explore scaling up this hybrid model using advanced architectures, such as cloud computing with MPI \citep{collier_distributed_2022}, high-performance platforms \citep{breitwieser_biodynamo_2022}, or database routing techniques \citep{solano_coupling_2013}, to apply the model across a broader range of infectious diseases and real-world scenarios. These approaches could address the computational demands of large agent populations, presenting a promising direction for further exploration.

Another area of exploration involves the effect of population sizes and model configurations. Stochastic effects in small populations can create discrepancies between deterministic compartmental models and stochastic ABMs \citep{mohd_revisiting_2022}. For infectious disease models, \citet{rahmandad_heterogeneity_2008} note that differences in model results are largely independent of population sizes ranging from 50 to 800, particularly when $R_0$ is high. Conversely, \citet{hunter_comparison_2018} observed a stronger correlation between population size and outcomes in compartmental models compared to ABMs. Notably, differences in time discretization between these models are frequently overlooked, contributing to variations in results \citep{ozmen_analyzing_2016}. In our experiments, population sizes within model components varied based on the assigned component weights. Future research should examine the effects of time discretization and parameter variation in hybrid models while isolating the influence of population size differences on model consistency.

\section{Conclusion}

In this paper, we addressed a significant research gap by introducing a novel hybrid modeling approach that spatially integrates a microscopic agent-based model (ABM) with a macroscopic compartmental ordinary differential equation (ODE) model. This topic has been scarcely explored in the limited literature on hybrid models, and most studies lack clear descriptions of model architecture and systematic investigations on result consistency. Our primary aim was to explore the feasibility of this hybrid model in reducing computational costs while maintaining accurate and detailed infection dynamics. We demonstrated how such coupling can be effectively achieved, highlighting the potential of spatial coupling in real-world epidemic modeling, advancing the understanding of hybrid systems, and setting a foundation for future research in this domain.

To evaluate the effectiveness of our approach, we utilized several performance metrics to detect coupling effects on the resulting infectious disease dynamics. Our experiment with two toy scenarios revealed that the outcomes of the hybrid model depend on the similarity between the integrated models. In both scenarios, the ODE component lacked the spatial resolution of the ABM component. However, in the first scenario, where both sub-models describe a homogeneous population, the hybrid model produced the same infectious disease dynamics as the reference model. In the second scenario, we introduced spatial heterogeneity within the ABM by enabling landscape-driven agent movement and localized clustering around an attraction point. In the corresponding hybrid model, disease dynamics were influenced by the location of the between-model interface relative to the attraction point.

To address this, we suggest a meta-population approach for the ODE-component, which could provide spatial resolution and heterogeneity without significantly increasing computational demands. We also acknowledge that the extreme variations in agent density in our toy examples may not be as pronounced in real-world applications, potentially mitigating concerns about model discrepancies.

To focus our analysis on the coupling effects, we intentionally simplified the model components, maintaining the standard susceptible, infectious, and recovered categories in both sub-models. This allowed us to evaluate the impact of spatial integration without the added complexity of more elaborate disease states or interactions. However, increasing the number of categories could further highlight the computational advantages of the hybrid approach compared to a purely microscopic approach. Future research will explore more realistic scenarios and model configurations to further validate and enhance our proposed hybrid model.

In conclusion, our spatial hybrid modeling approach presents a promising avenue for studying infectious diseases and complex systems. It offers opportunities to refine model accuracy and optimize computational efficiency, paving the way for more sophisticated and practical applications in future research endeavors.

\section{Implementation}
The model is implemented in Python v3.9 \citep{python_software_foundation_python_2020}. The project repository is available on GitHub (\url{git.zib.de/ibostanc/hybrid_abm_ode}), including the model code and jupyter notebooks for a tutorial and the presented experiments. The model involves three primary class objects: \textit{persons} creates an agent, \textit{model\_ode} solves the ODEs, and \textit{hybrid\_model} creates the ABM and couples it to the ODE. 

\section{Acknowledgements}

This work was funded by the German Ministry of research and education (BMBF) (project MODUS-Covid, grant number 031L0302C), and under Germany’s Excellence Strategy (MATH+ -The Berlin Mathematics Research Center, grant number EXC-2046/1, project no. 390685689, subproject EF45-4). The authors extend their gratitude to Kai Nagel and Kristina Maier for their valuable insights and comments on an early draft of this manuscript.

\bibliography{references}

\begin{thebibliography}{44}
\providecommand{\natexlab}[1]{#1}
\providecommand{\url}[1]{\texttt{#1}}
\expandafter\ifx\csname urlstyle\endcsname\relax
  \providecommand{\doi}[1]{doi: #1}\else
  \providecommand{\doi}{doi: \begingroup \urlstyle{rm}\Url}\fi

\bibitem[{Lancet Comission}(2024)]{lancet_comission_how_2024}
{Lancet Comission}.
\newblock How modelling can better support public health policy making: the {Lancet} {Commission} on {Strengthening} the {Use} of {Epidemiological} {Modelling} of {Emerging} and {Pandemic} {Infectious} {Diseases}.
\newblock \emph{The Lancet}, 403\penalty0 (10429):\penalty0 789--791, March 2024.
\newblock ISSN 01406736.
\newblock \doi{10.1016/S0140-6736(23)02758-7}.
\newblock URL \url{https://linkinghub.elsevier.com/retrieve/pii/S0140673623027587}.

\bibitem[Skrip and Townsend(2019)]{krause_modeling_2019}
Laura~A. Skrip and Jeffrey~P. Townsend.
\newblock Modeling {Approaches} {Toward} {Understanding} {Infectious} {Disease} {Transmission}.
\newblock In Peter~J. Krause, Paula~B. Kavathas, and Nancy~H. Ruddle, editors, \emph{Immunoepidemiology}, pages 227--243. Springer International Publishing, Cham, 2019.
\newblock ISBN 978-3-030-25552-7 978-3-030-25553-4.
\newblock \doi{10.1007/978-3-030-25553-4_14}.
\newblock URL \url{http://link.springer.com/10.1007/978-3-030-25553-4_14}.

\bibitem[Becker et~al.(2021)Becker, Grantz, Hegde, Bérubé, Cummings, and Wesolowski]{becker_development_2021}
Alexander~D Becker, Kyra~H Grantz, Sonia~T Hegde, Sophie Bérubé, Derek A~T Cummings, and Amy Wesolowski.
\newblock Development and dissemination of infectious disease dynamic transmission models during the {COVID}-19 pandemic: what can we learn from other pathogens and how can we move forward?
\newblock \emph{The Lancet Digital Health}, 3\penalty0 (1):\penalty0 e41--e50, January 2021.
\newblock ISSN 25897500.
\newblock \doi{10.1016/S2589-7500(20)30268-5}.
\newblock URL \url{https://linkinghub.elsevier.com/retrieve/pii/S2589750020302685}.

\bibitem[Hethcote(2000)]{hethcote_mathematics_2000}
Herbert~W. Hethcote.
\newblock The {Mathematics} of {Infectious} {Diseases}.
\newblock \emph{SIAM Review}, 42\penalty0 (4):\penalty0 599--653, 2000.
\newblock \doi{10.1137/S0036144500371907}.
\newblock URL \url{https://doi.org/10.1137/S0036144500371907}.
\newblock \_eprint: https://doi.org/10.1137/S0036144500371907.

\bibitem[Brauer et~al.(2019)Brauer, Castillo-Chavez, and Feng]{brauer_mathematical_2019}
Fred Brauer, Carlos Castillo-Chavez, and Zhilan Feng.
\newblock \emph{Mathematical {Models} in {Epidemiology}}, volume~69 of \emph{Texts in {Applied} {Mathematics}}.
\newblock Springer New York, New York, NY, 2019.
\newblock ISBN 978-1-4939-9826-5 978-1-4939-9828-9.
\newblock \doi{10.1007/978-1-4939-9828-9}.
\newblock URL \url{https://link.springer.com/10.1007/978-1-4939-9828-9}.

\bibitem[Duan et~al.(2015)Duan, Fan, Zhang, Guo, and Qiu]{duan_mathematical_2015}
Wei Duan, Zongchen Fan, Peng Zhang, Gang Guo, and Xiaogang Qiu.
\newblock Mathematical and computational approaches to epidemic modeling: a comprehensive review.
\newblock \emph{Frontiers of Computer Science}, 9\penalty0 (5):\penalty0 806--826, October 2015.
\newblock ISSN 2095-2228, 2095-2236.
\newblock \doi{10.1007/s11704-014-3369-2}.
\newblock URL \url{http://link.springer.com/10.1007/s11704-014-3369-2}.

\bibitem[Wulkow et~al.(2021)Wulkow, Conrad, Djurdjevac~Conrad, Müller, Nagel, and Schütte]{wulkow_prediction_2021}
Hanna Wulkow, Tim O.~F. Conrad, Nataša Djurdjevac~Conrad, Sebastian~A. Müller, Kai Nagel, and Christof Schütte.
\newblock Prediction of {Covid}-19 spreading and optimal coordination of counter-measures: {From} microscopic to macroscopic models to {Pareto} fronts.
\newblock \emph{PLOS ONE}, 16\penalty0 (4):\penalty0 e0249676, April 2021.
\newblock ISSN 1932-6203.
\newblock \doi{10.1371/journal.pone.0249676}.
\newblock URL \url{https://dx.plos.org/10.1371/journal.pone.0249676}.

\bibitem[Chang et~al.(2021)Chang, Pierson, Koh, Gerardin, Redbird, Grusky, and Leskovec]{chang_mobility_2021}
Serina Chang, Emma Pierson, Pang~Wei Koh, Jaline Gerardin, Beth Redbird, David Grusky, and Jure Leskovec.
\newblock Mobility network models of {COVID}-19 explain inequities and inform reopening.
\newblock \emph{Nature}, 589\penalty0 (7840):\penalty0 82--87, January 2021.
\newblock ISSN 0028-0836, 1476-4687.
\newblock \doi{10.1038/s41586-020-2923-3}.
\newblock URL \url{http://www.nature.com/articles/s41586-020-2923-3}.

\bibitem[Calvetti et~al.(2020)Calvetti, Hoover, Rose, and Somersalo]{calvetti_metapopulation_2020}
Daniela Calvetti, Alexander~P. Hoover, Johnie Rose, and Erkki Somersalo.
\newblock Metapopulation {Network} {Models} for {Understanding}, {Predicting}, and {Managing} the {Coronavirus} {Disease} {COVID}-19.
\newblock \emph{Frontiers in Physics}, 8:\penalty0 261, June 2020.
\newblock ISSN 2296-424X.
\newblock \doi{10.3389/fphy.2020.00261}.
\newblock URL \url{https://www.frontiersin.org/article/10.3389/fphy.2020.00261/full}.

\bibitem[Winkelmann et~al.(2021)Winkelmann, Zonker, Schütte, and Conrad]{winkelmann_mathematical_2021}
Stefanie Winkelmann, Johannes Zonker, Christof Schütte, and Nataša~Djurdjevac Conrad.
\newblock Mathematical modeling of spatio-temporal population dynamics and application to epidemic spreading.
\newblock \emph{Mathematical Biosciences}, 336:\penalty0 108619, June 2021.
\newblock ISSN 00255564.
\newblock \doi{10.1016/j.mbs.2021.108619}.
\newblock URL \url{https://linkinghub.elsevier.com/retrieve/pii/S0025556421000614}.

\bibitem[Gao et~al.(2022)Gao, Dai, Wang, Perra, and Wang]{gao_epidemic_2022}
Shupeng Gao, Xiangfeng Dai, Lin Wang, Nicola Perra, and Zhen Wang.
\newblock Epidemic {Spreading} in {Metapopulation} {Networks} {Coupled} {With} {Awareness} {Propagation}.
\newblock \emph{IEEE Transactions on Cybernetics}, pages 1--13, 2022.
\newblock ISSN 2168-2267, 2168-2275.
\newblock \doi{10.1109/TCYB.2022.3198732}.
\newblock URL \url{https://ieeexplore.ieee.org/document/9875017/}.

\bibitem[Railsback and Grimm(2019)]{railsback_agent-based_2019}
Steven~F. Railsback and Volker Grimm.
\newblock \emph{Agent-based and individual-based modeling: a practical introduction}.
\newblock Princeton University Press, Princeton ; Oxford, second edition edition, 2019.
\newblock ISBN 978-0-691-19082-2 978-0-691-19083-9.
\newblock OCLC: on1051137181.

\bibitem[Bruch and Atwell(2015)]{bruch_agent-based_2015}
Elizabeth Bruch and Jon Atwell.
\newblock Agent-{Based} {Models} in {Empirical} {Social} {Research}.
\newblock \emph{Sociological Methods \& Research}, 44\penalty0 (2):\penalty0 186--221, May 2015.
\newblock ISSN 0049-1241, 1552-8294.
\newblock \doi{10.1177/0049124113506405}.
\newblock URL \url{http://journals.sagepub.com/doi/10.1177/0049124113506405}.

\bibitem[Kerr et~al.(2021)Kerr, Stuart, Mistry, Abeysuriya, Rosenfeld, Hart, Núñez, Cohen, Selvaraj, Hagedorn, George, Jastrzębski, Izzo, Fowler, Palmer, Delport, Scott, Kelly, Bennette, Wagner, Chang, Oron, Wenger, Panovska-Griffiths, Famulare, and Klein]{kerr_covasim_2021}
Cliff~C. Kerr, Robyn~M. Stuart, Dina Mistry, Romesh~G. Abeysuriya, Katherine Rosenfeld, Gregory~R. Hart, Rafael~C. Núñez, Jamie~A. Cohen, Prashanth Selvaraj, Brittany Hagedorn, Lauren George, Michał Jastrzębski, Amanda~S. Izzo, Greer Fowler, Anna Palmer, Dominic Delport, Nick Scott, Sherrie~L. Kelly, Caroline~S. Bennette, Bradley~G. Wagner, Stewart~T. Chang, Assaf~P. Oron, Edward~A. Wenger, Jasmina Panovska-Griffiths, Michael Famulare, and Daniel~J. Klein.
\newblock Covasim: {An} agent-based model of {COVID}-19 dynamics and interventions.
\newblock \emph{PLOS Computational Biology}, 17\penalty0 (7):\penalty0 e1009149, July 2021.
\newblock ISSN 1553-7358.
\newblock \doi{10.1371/journal.pcbi.1009149}.
\newblock URL \url{https://dx.plos.org/10.1371/journal.pcbi.1009149}.

\bibitem[Arifin et~al.(2015)Arifin, Arifin, Pitts, Rahman, Nowreen, Madey, and Collins]{arifin_landscape_2015}
S.~M.~Niaz Arifin, Rumana~Reaz Arifin, Dilkushi De~Alwis Pitts, M.~Sohel Rahman, Sara Nowreen, Gregory~R. Madey, and Frank~H. Collins.
\newblock Landscape {Epidemiology} {Modeling} {Using} an {Agent}-{Based} {Model} and a {Geographic} {Information} {System}.
\newblock \emph{Land}, 4\penalty0 (2):\penalty0 378--412, 2015.
\newblock ISSN 2073-445X.
\newblock \doi{10.3390/land4020378}.
\newblock URL \url{https://www.mdpi.com/2073-445X/4/2/378}.

\bibitem[Müller et~al.(2021)Müller, Balmer, Charlton, Ewert, Neumann, Rakow, Schlenther, and Nagel]{muller_predicting_2021}
Sebastian~A. Müller, Michael Balmer, William Charlton, Ricardo Ewert, Andreas Neumann, Christian Rakow, Tilmann Schlenther, and Kai Nagel.
\newblock Predicting the effects of {COVID}-19 related interventions in urban settings by combining activity-based modelling, agent-based simulation, and mobile phone data.
\newblock \emph{PLOS ONE}, 16\penalty0 (10):\penalty0 e0259037, October 2021.
\newblock ISSN 1932-6203.
\newblock \doi{10.1371/journal.pone.0259037}.
\newblock URL \url{https://dx.plos.org/10.1371/journal.pone.0259037}.

\bibitem[Merler et~al.(2015)Merler, Ajelli, Fumanelli, Gomes, Piontti, Rossi, Chao, Longini, Halloran, and Vespignani]{merler_spatiotemporal_2015}
Stefano Merler, Marco Ajelli, Laura Fumanelli, Marcelo F~C Gomes, Ana Pastore~Y Piontti, Luca Rossi, Dennis~L Chao, Ira~M Longini, M~Elizabeth Halloran, and Alessandro Vespignani.
\newblock Spatiotemporal spread of the 2014 outbreak of {Ebola} virus disease in {Liberia} and the effectiveness of non-pharmaceutical interventions: a computational modelling analysis.
\newblock \emph{The Lancet Infectious Diseases}, 15\penalty0 (2):\penalty0 204--211, February 2015.
\newblock ISSN 14733099.
\newblock \doi{10.1016/S1473-3099(14)71074-6}.
\newblock URL \url{https://linkinghub.elsevier.com/retrieve/pii/S1473309914710746}.

\bibitem[Brailsford et~al.(2019)Brailsford, Eldabi, Kunc, Mustafee, and Osorio]{brailsford_hybrid_2019}
Sally~C. Brailsford, Tillal Eldabi, Martin Kunc, Navonil Mustafee, and Andres~F. Osorio.
\newblock Hybrid simulation modelling in operational research: {A} state-of-the-art review.
\newblock \emph{European Journal of Operational Research}, 278\penalty0 (3):\penalty0 721--737, November 2019.
\newblock ISSN 03772217.
\newblock \doi{10.1016/j.ejor.2018.10.025}.
\newblock URL \url{https://linkinghub.elsevier.com/retrieve/pii/S0377221718308786}.

\bibitem[Frias-Martinez et~al.(2011)Frias-Martinez, Williamson, and Frias-Martinez]{frias-martinez_agent-based_2011}
Enrique Frias-Martinez, Graham Williamson, and Vanessa Frias-Martinez.
\newblock An {Agent}-{Based} {Model} of {Epidemic} {Spread} {Using} {Human} {Mobility} and {Social} {Network} {Information}.
\newblock In \emph{2011 {IEEE} {Third} {International} {Conference} on {Privacy}, {Security}, {Risk} and {Trust} and 2011 {IEEE} {Third} {International} {Conference} on {Social} {Computing}}, pages 57--64, Boston, MA, October 2011. IEEE.
\newblock ISBN 978-1-4577-1931-8.
\newblock \doi{10.1109/PASSAT/SocialCom.2011.142}.
\newblock URL \url{https://ieeexplore.ieee.org/document/6113095/}.

\bibitem[Mokhtari et~al.(2021)Mokhtari, Mineo, Kriseman, Kremer, Neal, and Larson]{mokhtari_multi-method_2021}
Amir Mokhtari, Cameron Mineo, Jeffrey Kriseman, Pedro Kremer, Lauren Neal, and John Larson.
\newblock A multi-method approach to modeling {COVID}-19 disease dynamics in the {United} {States}.
\newblock \emph{Scientific Reports}, 11\penalty0 (1):\penalty0 12426, June 2021.
\newblock ISSN 2045-2322.
\newblock \doi{10.1038/s41598-021-92000-w}.
\newblock URL \url{https://www.nature.com/articles/s41598-021-92000-w}.

\bibitem[Özmen et~al.(2016)Özmen, Nutaro, Pullum, and Ramanathan]{ozmen_analyzing_2016}
Özgür Özmen, James~J Nutaro, Laura~L Pullum, and Arvind Ramanathan.
\newblock Analyzing the impact of modeling choices and assumptions in compartmental epidemiological models.
\newblock \emph{SIMULATION}, 92\penalty0 (5):\penalty0 459--472, May 2016.
\newblock ISSN 0037-5497, 1741-3133.
\newblock \doi{10.1177/0037549716640877}.
\newblock URL \url{http://journals.sagepub.com/doi/10.1177/0037549716640877}.

\bibitem[Hunter et~al.(2018)Hunter, Mac~Mamee, and Kelleher]{hunter_comparison_2018}
Elizabeth Hunter, Brain Mac~Mamee, and John~D Kelleher.
\newblock A {Comparison} of {Agent}-{Based} {Models} and {Equation} {Based} {Models} for {Infectious} {Disease} {Epidemiology}.
\newblock In \emph{26th {AIAI} {Irish} {Conference} on {Artificial} {Intelligence} and {Cognitive} {Science}}, Dublin, 2018.
\newblock \doi{10.21427/RTQ2-HS52}.
\newblock URL \url{https://arrow.dit.ie/scschcomart/71/}.
\newblock Publisher: Technological University of Dublin.

\bibitem[Bobashev et~al.(2007)Bobashev, Goedecke, {Feng Yu}, and Epstein]{bobashev_hybrid_2007}
Georgiy~V. Bobashev, D.~Michael Goedecke, {Feng Yu}, and Joshua~M. Epstein.
\newblock A {Hybrid} {Epidemic} {Model}: {Combining} {The} {Advantages} {Of} {Agent}-{Based} {And} {Equation}-{Based} {Approaches}.
\newblock In \emph{2007 {Winter} {Simulation} {Conference}}, pages 1532--1537, Washington, DC, USA, 2007. IEEE.
\newblock ISBN 978-1-4244-1305-8.
\newblock \doi{10.1109/WSC.2007.4419767}.
\newblock URL \url{http://ieeexplore.ieee.org/document/4419767/}.

\bibitem[Hunter et~al.(2020)Hunter, Mac~Namee, and Kelleher]{hunter_hybrid_2020}
Elizabeth Hunter, Brian Mac~Namee, and John Kelleher.
\newblock A {Hybrid} {Agent}-{Based} and {Equation} {Based} {Model} for the {Spread} of {Infectious} {Diseases}.
\newblock \emph{Journal of Artificial Societies and Social Simulation}, 23\penalty0 (4):\penalty0 14, 2020.
\newblock ISSN 1460-7425.
\newblock \doi{10.18564/jasss.4421}.
\newblock URL \url{http://jasss.soc.surrey.ac.uk/23/4/14.html}.

\bibitem[Nghi et~al.(2016)Nghi, Nguyen-Huu, Grignard, Huynh, and Drogoul]{vinh_toward_2016}
Huynh~Quang Nghi, Tri Nguyen-Huu, Arnaud Grignard, Hiep~Xuan Huynh, and Alexis Drogoul.
\newblock Toward an {Agent}-{Based} and {Equation}-{Based} {Coupling} {Framework}.
\newblock In Phan~Cong Vinh and Leonard Barolli, editors, \emph{Nature of {Computation} and {Communication}}, volume 168, pages 311--324. Springer International Publishing, Cham, 2016.
\newblock ISBN 978-3-319-46908-9 978-3-319-46909-6.
\newblock \doi{10.1007/978-3-319-46909-6_28}.
\newblock URL \url{http://link.springer.com/10.1007/978-3-319-46909-6_28}.
\newblock Series Title: Lecture Notes of the Institute for Computer Sciences, Social Informatics and Telecommunications Engineering.

\bibitem[Nguyen et~al.(2022)Nguyen, Megiddo, and Howick]{nguyen_hybrid_2022}
Le~Khanh~Ngan Nguyen, Itamar Megiddo, and Susan Howick.
\newblock Hybrid simulation modelling of networks of heterogeneous care homes and the inter-facility spread of {Covid}-19 by sharing staff.
\newblock \emph{PLOS Computational Biology}, 18\penalty0 (1):\penalty0 e1009780, January 2022.
\newblock ISSN 1553-7358.
\newblock \doi{10.1371/journal.pcbi.1009780}.
\newblock URL \url{https://dx.plos.org/10.1371/journal.pcbi.1009780}.

\bibitem[Bradhurst et~al.(2016)Bradhurst, Roche, East, Kwan, and Garner]{bradhurst_improving_2016}
R.A. Bradhurst, S.E. Roche, I.J. East, P.~Kwan, and M.G. Garner.
\newblock Improving the computational efficiency of an agent-based spatiotemporal model of livestock disease spread and control.
\newblock \emph{Environmental Modelling \& Software}, 77:\penalty0 1--12, March 2016.
\newblock ISSN 13648152.
\newblock \doi{10.1016/j.envsoft.2015.11.015}.
\newblock URL \url{https://linkinghub.elsevier.com/retrieve/pii/S1364815215301043}.

\bibitem[Banos et~al.(2015)Banos, Corson, Gaudou, Laperrière, and Coyrehourcq]{banos_importance_2015}
Arnaud Banos, Nathalie Corson, Benoit Gaudou, Vincent Laperrière, and Sébastien Coyrehourcq.
\newblock The {Importance} of {Being} {Hybrid} for {Spatial} {Epidemic} {Models}:{A} {Multi}-{Scale} {Approach}.
\newblock \emph{Systems}, 3\penalty0 (4):\penalty0 309--329, November 2015.
\newblock ISSN 2079-8954.
\newblock \doi{10.3390/systems3040309}.
\newblock URL \url{http://www.mdpi.com/2079-8954/3/4/309}.

\bibitem[Marilleau et~al.(2018)Marilleau, Lang, and Giraudoux]{marilleau_coupling_2018}
Nicolas Marilleau, Christophe Lang, and Patrick Giraudoux.
\newblock Coupling agent-based with equation-based models to study spatially explicit megapopulation dynamics.
\newblock \emph{Ecological Modelling}, 384:\penalty0 34--42, September 2018.
\newblock ISSN 03043800.
\newblock \doi{10.1016/j.ecolmodel.2018.06.011}.
\newblock URL \url{https://linkinghub.elsevier.com/retrieve/pii/S0304380018302163}.

\bibitem[Gustafsson and Sternad(2016)]{gustafsson_guide_2016}
Leif Gustafsson and Mikael Sternad.
\newblock A {Guide} to {Population} {Modelling} for {Simulation}.
\newblock \emph{Open Journal of Modelling and Simulation}, 04\penalty0 (02):\penalty0 55--92, 2016.
\newblock ISSN 2327-4018, 2327-4026.
\newblock \doi{10.4236/ojmsi.2016.42007}.
\newblock URL \url{http://www.scirp.org/journal/doi.aspx?DOI=10.4236/ojmsi.2016.42007}.

\bibitem[Sewall et~al.(2011)Sewall, Wilkie, and Lin]{sewall_interactive_2011}
Jason Sewall, David Wilkie, and Ming~C. Lin.
\newblock Interactive hybrid simulation of large-scale traffic.
\newblock In \emph{Proceedings of the 2011 {SIGGRAPH} {Asia} {Conference}}, pages 1--12, Hong Kong China, December 2011. ACM.
\newblock ISBN 978-1-4503-0807-6.
\newblock \doi{10.1145/2024156.2024169}.
\newblock URL \url{https://dl.acm.org/doi/10.1145/2024156.2024169}.

\bibitem[Solano et~al.(2013)Solano, Morris, and Bobashev]{solano_coupling_2013}
Eric Solano, Robert Morris, and Georgiy Bobashev.
\newblock Coupling models by routing communication through a database.
\newblock Technical report, RTI Press, September 2013.
\newblock URL \url{http://www.rti.org/publication/coupling-models-routing-communication-through-database}.

\bibitem[Breitwieser et~al.(2022)Breitwieser, Hesam, De~Montigny, Vavourakis, Iosif, Jennings, Kaiser, Manca, Di~Meglio, Al-Ars, Rademakers, Mutlu, and Bauer]{breitwieser_biodynamo_2022}
Lukas Breitwieser, Ahmad Hesam, Jean De~Montigny, Vasileios Vavourakis, Alexandros Iosif, Jack Jennings, Marcus Kaiser, Marco Manca, Alberto Di~Meglio, Zaid Al-Ars, Fons Rademakers, Onur Mutlu, and Roman Bauer.
\newblock {BioDynaMo}: a modular platform for high-performance agent-based simulation.
\newblock \emph{Bioinformatics}, 38\penalty0 (2):\penalty0 453--460, January 2022.
\newblock ISSN 1367-4803, 1367-4811.
\newblock \doi{10.1093/bioinformatics/btab649}.
\newblock URL \url{https://academic.oup.com/bioinformatics/article/38/2/453/6371176}.

\bibitem[Jones and Su(2015)]{jones_dose-response_2015}
Rachael~M Jones and Yu-Min Su.
\newblock Dose-response models for selected respiratory infectious agents: {Bordetella} pertussis, group a {Streptococcus}, rhinovirus and respiratory syncytial virus.
\newblock \emph{BMC Infectious Diseases}, 15\penalty0 (1):\penalty0 90, December 2015.
\newblock ISSN 1471-2334.
\newblock \doi{10.1186/s12879-015-0832-0}.
\newblock URL \url{https://bmcinfectdis.biomedcentral.com/articles/10.1186/s12879-015-0832-0}.

\bibitem[Aganovic et~al.(2023)Aganovic, Cao, Kurnitski, and Wargocki]{aganovic_new_2023}
Amar Aganovic, Guangyu Cao, Jarek Kurnitski, and Pawel Wargocki.
\newblock New dose-response model and {SARS}-{CoV}-2 quanta emission rates for calculating the long-range airborne infection risk.
\newblock \emph{Building and Environment}, 228:\penalty0 109924, January 2023.
\newblock ISSN 03601323.
\newblock \doi{10.1016/j.buildenv.2022.109924}.
\newblock URL \url{https://linkinghub.elsevier.com/retrieve/pii/S0360132322011544}.

\bibitem[Djurdjevac~Conrad et~al.(2018)Djurdjevac~Conrad, Helfmann, Zonker, Winkelmann, and Schütte]{djurdjevac_conrad_human_2018}
Nataša Djurdjevac~Conrad, Luzie Helfmann, Johannes Zonker, Stefanie Winkelmann, and Christof Schütte.
\newblock Human mobility and innovation spreading in ancient times: a stochastic agent-based simulation approach.
\newblock \emph{EPJ Data Science}, 7\penalty0 (1):\penalty0 24, December 2018.
\newblock ISSN 2193-1127.
\newblock \doi{10.1140/epjds/s13688-018-0153-9}.
\newblock URL \url{https://epjdatascience.springeropen.com/articles/10.1140/epjds/s13688-018-0153-9}.

\bibitem[Marshall and Duthie(2022)]{marshall_abmanimalmovement_2022}
BM~Marshall and AB~Duthie.
\newblock {abmAnimalMovement}: {An} {R} package for simulating animal movement using an agent-based model [version 1; peer review: 2 approved with reservations].
\newblock \emph{F1000Research}, 11\penalty0 (1182), 2022.
\newblock \doi{10.12688/f1000research.124810.1}.

\bibitem[Kermack et~al.(1927)Kermack, McKendrick, and Walker]{kermack_contribution_1927}
William~Ogilvy Kermack, A.~G. McKendrick, and Gilbert~Thomas Walker.
\newblock A contribution to the mathematical theory of epidemics.
\newblock \emph{Proceedings of the Royal Society of London. Series A, Containing Papers of a Mathematical and Physical Character}, 115\penalty0 (772):\penalty0 700--721, 1927.
\newblock \doi{10.1098/rspa.1927.0118}.
\newblock URL \url{https://royalsocietypublishing.org/doi/abs/10.1098/rspa.1927.0118}.
\newblock \_eprint: https://royalsocietypublishing.org/doi/pdf/10.1098/rspa.1927.0118.

\bibitem[Susandi et~al.(2021)Susandi, Taufik, Aditiawati, and Viridi]{susandi_relation_2021}
Armi Susandi, Intan Taufik, Pingkan Aditiawati, and Sparisoma Viridi.
\newblock The relation between agent-based model and susceptible-infected-recovered model for spread of disease.
\newblock \emph{AIP Conference Proceedings}, 2320\penalty0 (1):\penalty0 050032, March 2021.
\newblock ISSN 0094-243X.
\newblock \doi{10.1063/5.0038221}.
\newblock URL \url{https://doi.org/10.1063/5.0038221}.
\newblock \_eprint: https://pubs.aip.org/aip/acp/article-pdf/doi/10.1063/5.0038221/14227268/050032\_1\_online.pdf.

\bibitem[Rahmandad and Sterman(2008)]{rahmandad_heterogeneity_2008}
Hazhir Rahmandad and John Sterman.
\newblock Heterogeneity and {Network} {Structure} in the {Dynamics} of {Diffusion}: {Comparing} {Agent}-{Based} and {Differential} {Equation} {Models}.
\newblock \emph{Management Science}, 54\penalty0 (5):\penalty0 998--1014, May 2008.
\newblock ISSN 0025-1909, 1526-5501.
\newblock \doi{10.1287/mnsc.1070.0787}.
\newblock URL \url{https://pubsonline.informs.org/doi/10.1287/mnsc.1070.0787}.

\bibitem[Virtanen et~al.(2020)Virtanen, Gommers, Oliphant, Haberland, Reddy, Cournapeau, Burovski, Peterson, Weckesser, Bright, and {others}]{virtanen_scipy_2020}
Pauli Virtanen, Ralf Gommers, Travis~E Oliphant, Matt Haberland, Tyler Reddy, David Cournapeau, Evgeni Burovski, Pearu Peterson, Warren Weckesser, Jonathan Bright, and {others}.
\newblock {SciPy} 1.0: fundamental algorithms for scientific computing in {Python}.
\newblock \emph{Nature methods}, 17\penalty0 (3):\penalty0 261--272, 2020.
\newblock Publisher: Nature Publishing Group.

\bibitem[Collier and Ozik(2022)]{collier_distributed_2022}
Nicholson Collier and Jonathan Ozik.
\newblock Distributed {Agent}-{Based} {Simulation} with {Repast4Py}.
\newblock In \emph{2022 {Winter} {Simulation} {Conference} ({WSC})}, pages 192--206, Singapore, December 2022. IEEE.
\newblock ISBN 978-1-66547-661-4.
\newblock \doi{10.1109/WSC57314.2022.10015389}.
\newblock URL \url{https://ieeexplore.ieee.org/document/10015389/}.

\bibitem[Mohd(2022)]{mohd_revisiting_2022}
Mohd~Hafiz Mohd.
\newblock Revisiting discrepancies between stochastic agent-based and deterministic models.
\newblock \emph{Community Ecology}, 23\penalty0 (3):\penalty0 453--468, October 2022.
\newblock ISSN 1585-8553, 1588-2756.
\newblock \doi{10.1007/s42974-022-00118-2}.
\newblock URL \url{https://link.springer.com/10.1007/s42974-022-00118-2}.

\bibitem[{Python Software Foundation}(2020)]{python_software_foundation_python_2020}
{Python Software Foundation}.
\newblock Python, 2020.
\newblock URL \url{https://www.python.org/}.
\newblock Archived at https://web.archive.org/web/20240718193634/ https://www.python.org/.

\end{thebibliography}
\appendix

\section{Appendix 1}

\begin{figure}[h]
 \begin{center}
  \begin{subfigure}[p]{0.33\textwidth}
    \includegraphics[width=\textwidth]{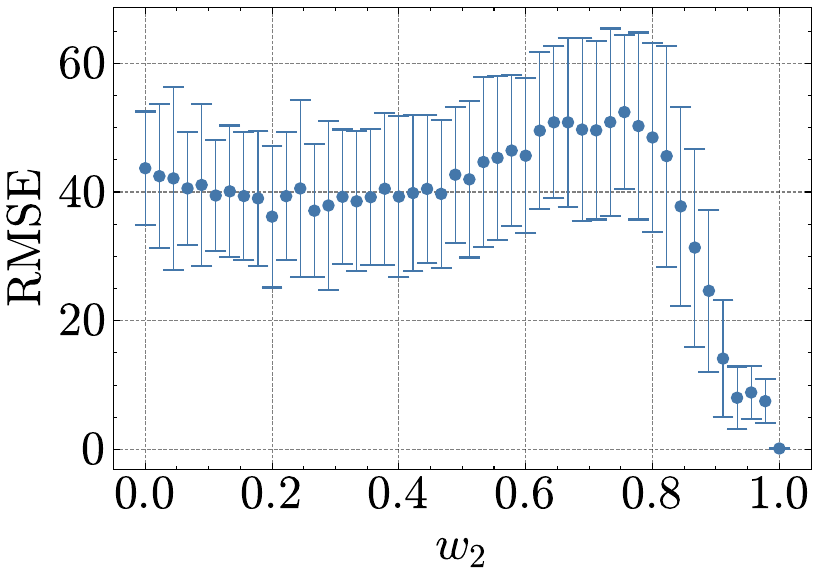}
    \caption{}
    \label{fig:rmsec}
  \end{subfigure}
  \begin{subfigure}[p]{0.33\textwidth}
    \includegraphics[width=\textwidth]{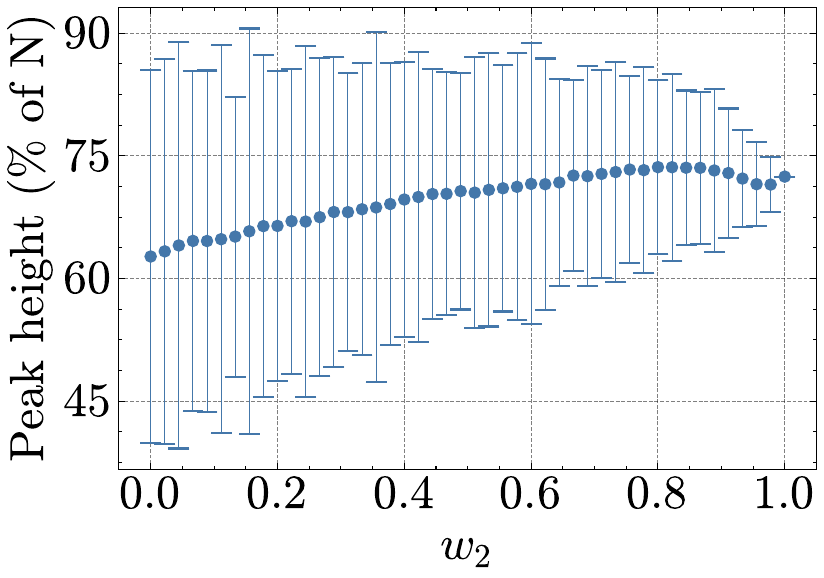}
    \caption{}
    \label{fig:max_infc}
  \end{subfigure}
  \begin{subfigure}[p]{0.33\textwidth}
    \includegraphics[width=\textwidth]{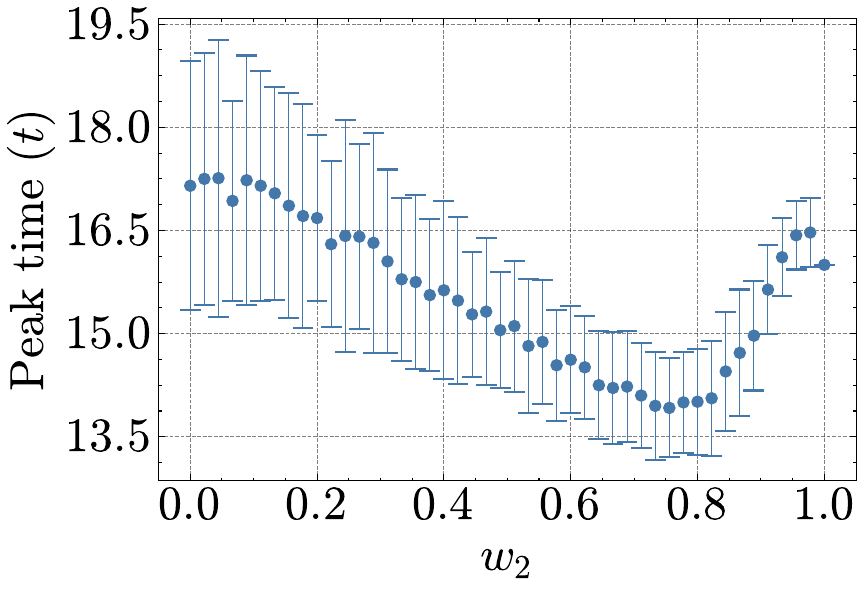}
    \caption{}
    \label{fig:t_peakc}
  \end{subfigure}
  \begin{subfigure}[p]{0.33\textwidth}
    \includegraphics[width=\textwidth]{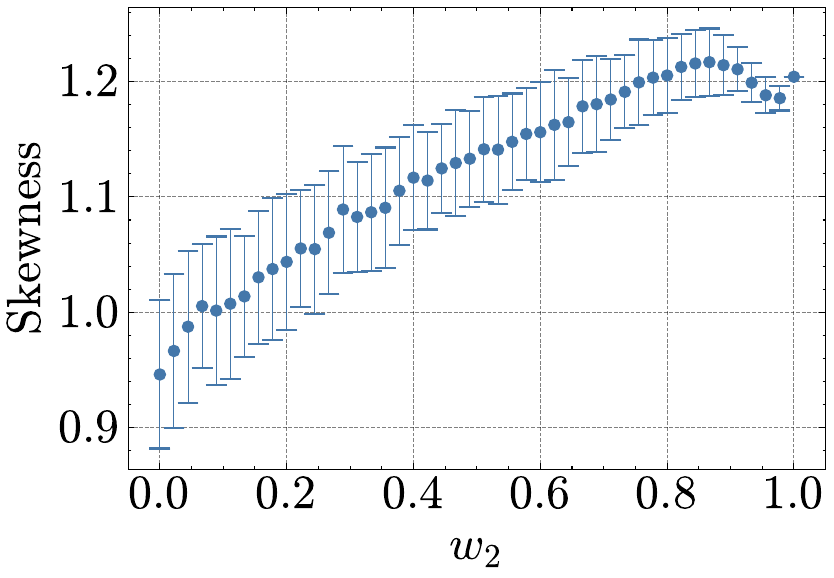}
    \caption{}
    \label{fig:skewnessc}
  \end{subfigure}
  \begin{subfigure}[p]{0.33\textwidth}
    \includegraphics[width=\textwidth]{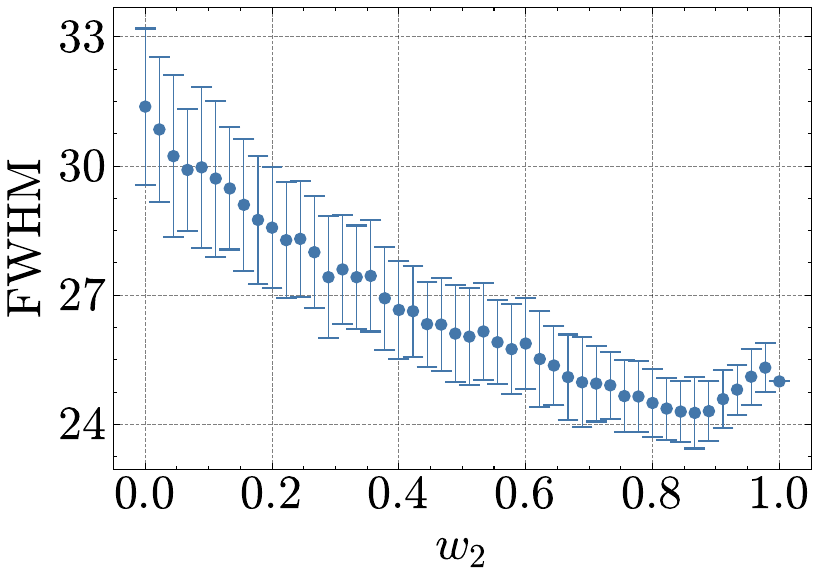}
    \caption{}
    \label{fig:fwhmc}
  \end{subfigure}
  \begin{subfigure}[p]{0.33\textwidth}
    \includegraphics[width=\textwidth]{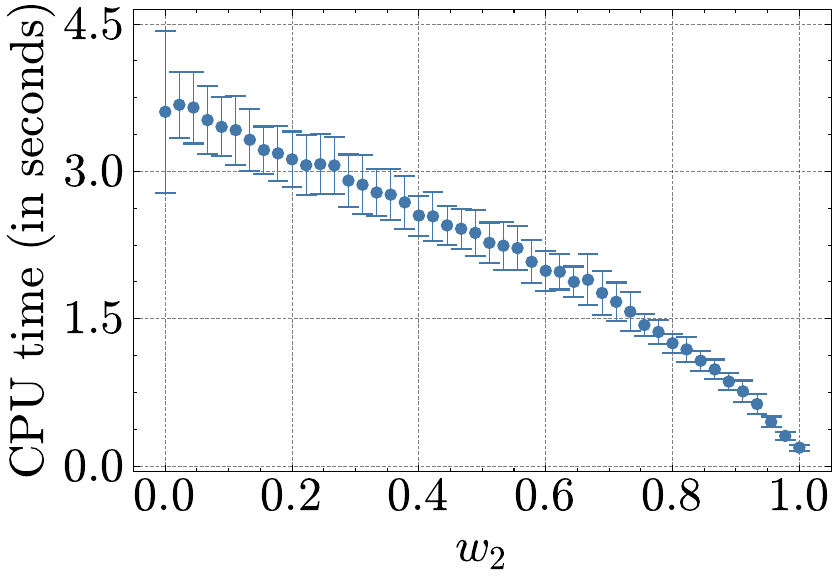}
    \caption{}
    \label{fig:cpu_timec}
  \end{subfigure}
   \end{center}
   \label{fig:hybc}
  \caption{Hybrid Model C: Average results for 100 independent realizations per level of M2 weight ($w_2$). Bars around points represent the standard deviation. (a) RMSE for the infectious time series between the reference model and the hybrid model. (b) Peak time of the infectious time series. (c) Peak height of the infectious time series. (d) Fisher-Pearson coefficient of skewness of the infectious time series. (e) FWHM of the infectious time series. (f) CPU time in seconds.}
\end{figure}

\begin{figure}[b]
 \begin{center}
  \begin{subfigure}[p]{0.36\textwidth}
    \includegraphics[width=\textwidth]{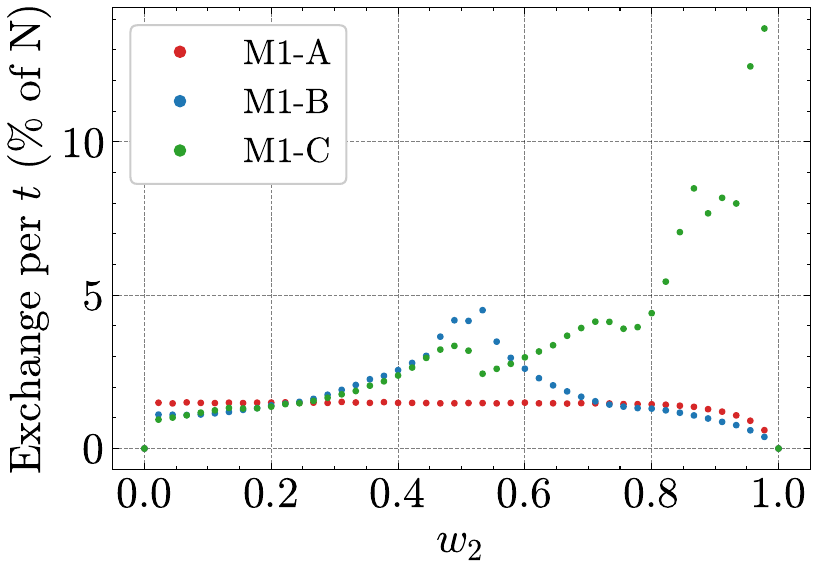}
    \caption{}
    \label{fig:mean_leave_abm_abc}
  \end{subfigure}
  \begin{subfigure}[p]{0.38\textwidth}
    \includegraphics[width=\textwidth]{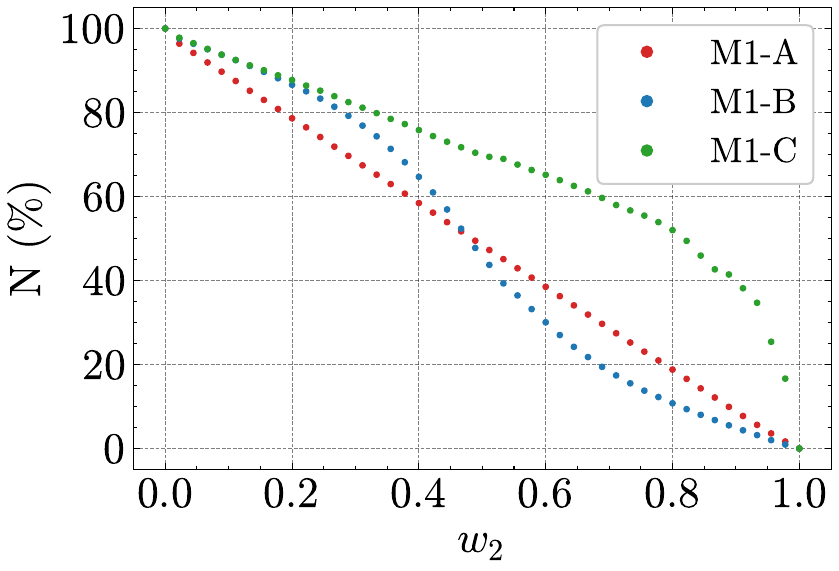}
    \caption{}
    \label{fig:mean_pop_abm_abc}
  \end{subfigure}
   \end{center}
  \caption{Changes in population metrics for Hybrid Model A, B, and C for different M2 weights ($w_2$). Results from 100 independent realizations. (a) Average percentage of population entering M2 per $t$. (b) Average percentage of total population in M1 per $t$.}
\end{figure}

Figures \ref{fig:rmsec} - \ref{fig:cpu_timec} show results for Hybrid Model C with the attraction point centered in M1. For example, at $w_2=0$ (pure ABM), the location matches Hybrid Model B at $(x=4.5, y=4.5)$, while at $w_2=0.5$, it is placed at $(x=2.25, y=4.5)$. Most results differ from Hybrid Model A and B. Instead of a steady or irregular progression towards the reference model, most metrics initially converge towards the reference model but exceed/precede the reference value around $w_2=0.8$ before converging again. Notably, the RMSE increases from $w_2=0.2$ to $w_2=0.74$ before sharply dropping. Given the relationship between the attraction point location and the population density within M1 (and the distribution of the total population between components), the fitted parameters of the pure ABM are likely not suitable across different conditions. Further, while the attraction point remains within M1, the distance to the interface shrinks with increasing $w_2$. As depicted in Figure \ref{fig:mean_leave_abm_abc}, this drives a steady rise in between-model interaction, contrasting with Hybrid Model B's peak around $w_2=0.5$. Additionally, Figure \ref{fig:mean_pop_abm_abc} shows that the majority of the total population is located within M1 when $w_2<0.85$. Thus, for Hybrid Model C, $w_2$ alone does not a adequately reflect the influence of M2 on overall population dynamics.

\end{document}